\documentclass[12pt]{article}
\usepackage{latexsym}
\usepackage{amsmath,amsfonts}
\usepackage{times}
\allowdisplaybreaks[4]
\catcode`\@=11

\newcount\hour
\newcount\minute
\newtoks\amorpm \hour=\time\divide\hour by 60\minute
=\time{\multiply\hour by 60 \global\advance\minute by-\hour}
\edef\standardtime{{\ifnum\hour<12 \global\amorpm={am}%
        \else\global\amorpm={pm}\advance\hour by-12 \fi
        \ifnum\hour=0 \hour=12 \fi
        \number\hour:\ifnum\minute<10
        0\fi\number\minute\the\amorpm}}
\edef\militarytime{\number\hour:\ifnum\minute<10
0\fi\number\minute}

\def\draftlabel#1{{\@bsphack\if@filesw {\let\thepage\relax
   \xdef\@gtempa{\write\@auxout{\string
      \newlabel{#1}{{\@currentlabel}{\thepage}}}}}\@gtempa
   \if@nobreak \ifvmode\nobreak\fi\fi\fi\@esphack}
        \gdef\@eqnlabel{#1}}
\def\@eqnlabel{}
\def\@vacuum{}
\def\marginnote#1{}
\def\draftmarginnote#1{\marginpar{\raggedright\scriptsize\tt#1}}
\def\draft{
        \pagestyle{plain}
        \overfullrule=2pt
        \oddsidemargin -.5truein
        \def\@oddhead{\sl \phantom{\today\quad\militarytime} \hfil
        \smash{\Large\sl DRAFT} \hfil \today\quad\militarytime}
        \let\@evenhead\@oddhead
        \let\label=\draftlabel
        \let\marginnote=\draftmarginnote
        \def\ps@empty{\let\@mkboth\@gobbletwo
        \def\@oddfoot{\hfil \smash{\Large\sl DRAFT} \hfil}
        \let\@evenfoot\@oddhead}
        \def\@eqnnum{(\theequation)\rlap{\kern\marginparsep\tt\@eqnlabel}%
        \global\let\@eqnlabel\@vacuum}  }

\newcommand{\rf}[1]{(\ref{#1})}
\renewcommand{\theequation}{\thesection.\arabic{equation}}
\renewcommand{\thefootnote}{\fnsymbol{footnote}}
\newcommand{\newsection}{    % Numeration of eqs. is automatic
\setcounter{equation}{0}\section}

\def\appendix#1{\addtocounter{section}{1}\setcounter{equation}{0}
\renewcommand{\thesection}{\Alph{section}}
\section*{Appendix \thesection\protect\indent \parbox[t]{11.15cm}{#1}}
\addcontentsline{toc}{section}{Appendix \thesection\ \ \ #1}}

\def\nline{\,\nabla\kern -0.7em\raise0.2ex\hbox{/}\,\,}
\def\yline{\,y\kern -0.47em /}
\def\aline{\,a\kern -0.49em /}
\def\parline{\,\partial\kern -0.55em /\,\,}

\newcommand{\Mo}{\mathbb{M}}
\newcommand{\No}{\mathbb{N}}
\newcommand{\Po}{\mathbb{P}}

\def\be{\begin{equation}}
\def\ee{\end{equation}}
\def\beq{\begin{eqnarray}}
\def\eeq{\end{eqnarray}}

\def\smpt{{\scriptscriptstyle [2]}}
\def\smp3{{\scriptscriptstyle [3]}}

\def\smpn{{\scriptscriptstyle [n]}}

\def\Thsm{{\scriptscriptstyle \rm Th}}
\def\scalarrm{{\scriptscriptstyle \rm scalar}}

\def\Jbf{{\bf J}}
\def\Kbf{{\bf K}}

\def\Mbf{{\bf M}}
\def\Pbf{{\bf P}}

\def\Xbf{{\bf X}}

\def\Hsm{{\scriptscriptstyle H}}
\def\Rsm{{\scriptscriptstyle R}}
\def\Lsm{{\scriptscriptstyle L}}

\def\asf{{\sf a}}
\def\bsf{{\sf b}}
\def\csf{{\sf c}}
\def\Nsf{{\sf N}}

\def\Jbf{{\bf J}}

\def\Mbf{{\bf M}}
\def\Pbf{{\bf P}}

\def\Xbf{{\bf X}}

\def\ibf{{\bf i}}
\def\iibf{{\bf ii}}
\def\iiibf{{\bf iii}}
\def\ivbf{{\bf iv}}
\def\vbf{{\bf v}}
\def\vibf{{\bf vi}}

\def\MM{{\cal M}}
\def\NN{{\cal N}}
\def\PP{{\cal P}}

\def\vph{{\vphantom{5pt}}}

\def\half{\frac{1}{2}}

\def\alphab{\bar{\alpha}}

\def\Cb{{\bar{C}}}

\def\Vb{{\bar{V}}}
\def\Ub{{\bar{U}}}
\def\Yb{{\bar{Y}}}
\def\ub{{\bar{u}}}

\def\phik{|\phi\rangle}

\def\irm{{\rm i}}

\def\dyn{{\rm dyn}}

\def\betach{\check{\beta}}

\begin{document}

%\draft

\begin{flushright}
FIAN-TD-2018-15 \ \ \ \ \ \ \  \\
arXiv: 1807.07542 V2
\end{flushright}

\vspace{1cm}

\begin{center}

{\Large \bf Light-cone gauge cubic interaction

\medskip
vertices for massless fields in AdS(4)}

\vspace{2.5cm}

R.R. Metsaev\footnote{ E-mail: metsaev@lpi.ru }

\vspace{1cm}

{\it Department of Theoretical Physics, P.N. Lebedev Physical
Institute, \\ Leninsky prospect 53,  Moscow 119991, Russia }

\vspace{3cm}

{\bf Abstract}

\end{center}

In the framework of light-cone formulation of relativistic dynamics, arbitrary spin massless fields propagating in the four-dimensional AdS space are studied. For such fields, the complete list of light-cone gauge cubic interaction vertices is obtained. Realization of relativistic symmetries on space of light-cone gauge massless AdS fields is also obtained. The light-cone gauge vertices for  massless AdS fields take simple form similar to the one for massless fields in the flat space.

\vspace{3cm}

Higher-spin fields; Light-cone gauge approach; Cubic interaction vertices.

\newpage
\renewcommand{\thefootnote}{\arabic{footnote}}
\setcounter{footnote}{0}

%%%%%%%%%%%%%%%%%%%%%%%%%%%%%%%%%%%%%%%%%%%%%%%%%%%%%%%%%%%%%%%%%
\newsection{ \large Introduction}
%%%%%%%%%%%%%%%%%%%%%%%%%%%%%%%%%%%%%%%%%%%%%%%%%%%%%%%%%%%%%%%%%

Light-cone gauge formulation of field dynamics in the flat space developed in Ref.\cite{Dirac:1949cp} has turned out to be successful for a study of many important problems of field/string theory.
Perhaps one of the attractive applications of the light-cone formalism is the construction of the light-cone gauge (super)string field theory in Refs.\cite{Kaku:1974zz,Green:1983hw}. The light-cone gauge superfield formulation of $\NN=4$ supersymmetric Yang-Mills theory was built in Refs.\cite{Brink:1982pd}, while the light-cone gauge superfield formulations of the supergravity theories in ten and eleven dimensions were studied in Refs.\cite{Green:1982tk,Metsaev:2004wv,Ananth:2005vg}.
Another attractive application of the light-cone formalism is the construction of
interaction vertices in the theory of higher-spin fields propagating in flat space
\cite{Bengtsson:1983pd}-\cite{Metsaev:2007rn}. Recent interesting applications of light-cone gauge formalism for the study of various dynamical system in flat space may be found in Refs.\cite{Conde:2016izb,Metsaev:2017cuz}.

Light-cone gauge formulation of field dynamics in AdS space was developed in Refs.\cite{Metsaev:1999ui,Metsaev:2003cu}. In Refs.\cite{Metsaev:1999ui,Metsaev:2003cu}, we studied {\it free} light-cone gauge fields propagating in AdS space. Our aim in this paper is to develop light-cone gauge formulation for {\it interacting} massless fields propagating in four dimensional AdS space.
We develop the systematic method for building interaction vertices for arbitrary spin massless AdS fields and use this method to find explicit expressions for cubic interaction vertices. We find the complete list of cubic interaction vertices for massless AdS fields.%
\footnote{ Full equations of motion for higher-spin gauge AdS field were obtained in Refs.\cite{Vasiliev:1990en}. Recent discussion of this theme may be found, e.g., in Refs.\cite{Gelfond:2017wrh}. Discussion of AKSZ action for higher-spin gauge field may be found in Ref.\cite{Boulanger:2011dd}.}
As is well known the light-cone approach has turned out to be useful for the study of interacting superstring theory. By analogy with this, we believe that the light-cone gauge interaction vertices obtained in this paper will be useful for better understanding  of higher-spin field theory.

This paper is organized as follows.

In Sec.\ref{sec-02}, we
introduce our notation and describe the light-cone gauge formulation of free arbitrary spin massless fields propagating in $AdS_4$ space.

In Sec.\ref{sec-03},  we start with the discussion of $n$-point interaction vertices for massless AdS fields. We find restrictions imposed on the $n$-point interaction vertices by the kinematical symmetries of the $so(3,2)$ algebra. After that we restrict our attention to cubic vertices. Using a particular choice of momentum variables which suit the cubic vertices, we provide the convenient form of restrictions imposed on the cubic vertices by the kinematical symmetries of the $so(3,2)$ algebra.

In Sec.\ref{sec-04}, we discuss restrictions imposed on the cubic vertices by dynamical symmetries of the $so(3,2)$ algebra. After that, requiring the light-cone locality and using field redefinitions, we find the complete list of equations which admits us to determine cubic vertices uniquely.

In Sec.\ref{sec-05}, we present our method for solving the complete list of equations for the cubic vertices, while in Sec.\ref{sec-06} we summarize our final results for the cubic interaction vertices. We discuss various representations for the cubic vertices.

In Sec.\ref{sec-07}, we summarize our conclusions and suggest directions for future research.

Notation, conventions, and various technical details are collected in  Appendices. Appendix A is devoted to the basic notation and conventions we use in this paper. In Appendices B and C, we discuss various technical details of our method for solving equations for the cubic vertices.

%%%%%%%%%%%%%%%%%%%%%%%%%%%%%%%%%%%%%%%%%%%%%%%%%%%%%%%%%%%%%%%%%%%%%%%%%%%%%%%
%%%%%%%%%%%%%%%%%%%%%%%%%%%%%%%%%%%%%%%%%%%%%%%%%%%%%%%%%%%%%%%%%%%%%%%%%%%%%%%
\newsection{ \large Free light-cone gauge arbitrary spin massless fields in $AdS_4$ space }\label{sec-02}
%%%%%%%%%%%%%%%%%%%%%%%%%%%%%%%%%%%%%%%%%%%%%%%%%%%%%%%%%%%%%%%%%%%%%%%%%%%%%%%
%%%%%%%%%%%%%%%%%%%%%%%%%%%%%%%%%%%%%%%%%%%%%%%%%%%%%%%%%%%%%%%%%%%%%%%%%%%%%%%

\noindent {\bf so(3,2) algebra in light-cone frame}. According to the idea in Ref.\cite{Dirac:1949cp}, the problem of finding a new dynamical system  amounts to the problem
of finding a new (light cone gauge) solution for commutation relations of a basic symmetry algebra.
For fields that propagate in the $AdS_4$ space, basic symmetries are associated with the   algebra $so(3,2)$. Light-cone gauge formulation of free fields propagating in AdS space was developed in Ref.\cite{Metsaev:1999ui}.
Using the light-cone gauge approach in Ref.\cite{Metsaev:1999ui}, we now discuss the light-cone gauge realization of the $so(3,2)$ algebra symmetries on space of free arbitrary spin massless fields propagating in $AdS_4$ space.

The $so(3,2)$ algebra is spanned by translation generators $P^\mu$, dilatation generator $D$, conformal boost generators $K^\mu$ and rotation generators $J^{\mu\nu}$ which are generators of
the $so(2,1)$ algebra. The commutation relations of the $so(3,2)$ algebra take the form
\beq
\label{25062018-man02-01} && {}[D,P^\mu]=-P^\mu\,, \hspace{2.3cm}  {}[P^\mu,J^{\nu\rho}] = \eta^{\mu\nu} P^\rho - \eta^{\mu\rho} P^\nu, \qquad
\nonumber\\
&& [D,K^\mu]=K^\mu\,, \hspace{2.5cm} [K^\mu,J^{\nu\rho}]=\eta^{\mu\nu}K^\rho - \eta^{\mu\rho} K^\nu,
\qquad
\\
&& [P^\mu,K^\nu]=\eta^{\mu\nu}D - J^{\mu\nu}\,, \qquad  [J^{\mu\nu},J^{\rho\sigma}]=\eta^{\nu\rho}J^{\mu\sigma}+3\hbox{ terms} \,,
\nonumber
\eeq
where $\eta^{\mu\nu}$ is the mostly positive flat metric tensor.  The vector indices of the $so(2,1)$ algebra take values $\mu,\nu,\rho,\sigma=0,1,2$. The generators $P^\mu$ and $K^\mu$ are considered to be hermitian, while the generators $D$ and $J^{\mu\nu}$ are assumed to be anti-hermitian.

We use the Poincar\'e parametrization of $AdS_4$ space,
\be \label{25062018-man02-02}
ds^2= \frac{R^2}{z^2}(- dx^0 dx^0 + dx^1dx^1 + dx^2 dx^2 + dz dz)\,.
\ee

\vspace{-0.2cm}
To discuss the light-cone formulation, we introduce, in place of
the coordinates $x^0$, $x^1$, $x^2$, $z$, the light-cone basis coordinates $x^+$, $x^-$, $x^1$, $z$, where the coordinates $x^\pm$ are defined as
\be \label{25062018-man02-03}
x^\pm \equiv \frac{1}{\sqrt{2}}(x^2  \pm x^0)\,.
\ee
The coordinate $x^+$ is considered as an evolution parameter. Use of the coordinates   \rf{25062018-man02-03} implies that the $so(2,1)$ algebra vector $X^\mu$ is decomposed as $X^+,X^-,X^1$, while a scalar product of the $so(2,1)$ algebra vectors $X^\mu$ and $Y^\mu$ is decomposed as
\be \label{25062018-man02-04}
\eta_{\mu\nu}X^\mu Y^\nu = X^+Y^- + X^-Y^+ + X^1 Y^1\,.
\ee
From decomposition \rf{25062018-man02-04}, we conclude, that in light-cone frame, non vanishing elements of the flat metric $\eta_{\mu\nu}$ and its inverse $\eta^{\mu\nu}$ are given by
\be \label{25062018-man02-05}
\eta_{+-} = 1,\quad \eta_{-+}=1,\quad \eta_{11}^\vph = 1,\qquad \eta^{+-} = 1,\quad \eta^{-+}=1,\quad \eta^{11} = 1.
\ee
Relations \rf{25062018-man02-05} imply  that the covariant and contravariant components of vectors are related as $X^+=X_-$, $X^-=X_+$, $X^1=X_1$.

In light-cone approach, generators of the $so(3,2)$ algebra are
separated into the following two groups:
\beq
\label{25062018-man02-06} &&
P^+,\quad
P^1,\quad
D,\quad
J^{+1},\quad
J^{+-},\quad
K^+,\quad
K^1, \hspace{2cm} \hbox{ kinematical generators};\qquad
\\
\label{25062018-man02-07} && P^-,\quad
J^{-1}, \quad K^- \hspace{6.8cm}
\hbox{ dynamical generators}.
\eeq
For $x^+=0$, in the field theoretical realization, kinematical generators \rf{25062018-man02-06} are
quadratic in fields%
\footnote{For arbitrary $x^+ \ne 0 $, kinematical generators \rf{25062018-man02-06} can be presented as
$G= G_1 + x^+ G_2 + x^+x^+G_3$, where a functional $G_1$ is quadratic in fields, while a functionals $G_2$, $G_3$
involve quadratic and higher order terms in fields.},
while, dynamical generators \rf{25062018-man02-07} involve quadratic and higher order terms in fields. In light-cone frame, commutation relations for the generators \rf{25062018-man02-06},\rf{25062018-man02-07} are
obtained from the ones in \rf{25062018-man02-01} by using the
non vanishing elements of the $\eta^{\mu\nu}$ given in \rf{25062018-man02-05}.

To provide a field theoretical realization for the generators of the $so(3,2)$ algebra on a space of arbitrary spin massless fields we exploit a light-cone gauge description of the fields.

\noindent {\bf Arbitrary spin-$s$ massless fields}. In light-cone gauge, physical degrees of freedom of a massless spin-$s$ field, $s>0$, propagating in $AdS_4$ space are described by two complex-valued fields $\phi_s$ and $\phi_{-s}$ that are hermitian conjugated to each other,
\be \label{27062018-man02-01}
\phi_s(x^+,x^-,x^1,z)\,, \qquad \phi_{-s}(x^+,x^-,x^1,z) \,, \qquad \phi_s^\dagger(x^+,x^-,x^1,z) = \phi_{-s}(x^+,x^-,x^1,z)\,,
\ee
while spin-0 field (scalar field)  can be described by real-valued field
\be \label{27062018-man02-02}
\phi_0(x^+,x^-,x^1,z)\,, \qquad \phi_0^\dagger(x^+,x^-,x^1,z) = \phi_0(x^+,x^-,x^1,z)\,.
\ee
We prefer to deal with fields obtained from the ones in \rf{27062018-man02-01},\rf{27062018-man02-02} by using the Fourier transform with respect to the coordinates $x^-$ and $x^1$,
\be \label{27062018-man02-03}
\phi_\lambda(x^+,x^-,x^1,z) = \int \frac{dp^1d\beta}{2\pi}\,\, e^{\irm (p^1x^1 + \beta x^-)} \phi_\lambda(x^+,\beta,p^1,z) \,,\qquad  \lambda =0,\pm s\,.
\ee
In other words, to discuss spin-$s$ field, $s>0$, and spin-0 field we use the following respective fields
\beq
\label{27062018-man02-04} && \phi_s(p,z)\,, \qquad \phi_{-s}(p,z) \,, \qquad \phi_s^\dagger(p,z) = \phi_{-s}(-p,z)\,,
\\
\label{27062018-man02-05} && \phi_0(p,z)\,, \hspace{3.5cm}  \phi_0^\dagger(p,z) = \phi_0(-p,z)\,,
\eeq
where, in \rf{27062018-man02-04},\rf{27062018-man02-05}, the argument $p$ stands for the momenta $p^1$, $\beta$ and the dependence on the evolution parameter $x^+$ is implicit.

In order to discuss the light-cone gauge formulation of a massless field in an easy-to-use form we introduce the creation operators $\alpha^\Rsm$, $\alpha^\Lsm$ and the respective annihilation operators $\alphab^\Lsm$, $\alphab^\Rsm$,
\be \label{27062018-man02-06}
[\alphab^\Rsm,\alpha^\Lsm] = 1\,,  \quad [\alphab^\Lsm,\alpha^\Rsm]=1\,, \quad \alphab^\Rsm |0\rangle = 0\,, \quad \alphab^\Lsm|0\rangle = 0\,, \quad \alpha^{\Rsm\dagger} = \alphab^\Lsm\,, \quad \alpha^{\Lsm\dagger} = \alphab^\Rsm\,.
\ee
Throughout this paper, the creation and annihilation operators will be referred to as oscillators.
Sometimes, we prefer to use oscillators with lower case indices defined by the relations
\be \label{27062018-man02-07}
\alpha_\Lsm = \alpha^\Rsm, \qquad \alpha_\Rsm = \alpha^\Lsm\,, \qquad
\alphab_\Lsm = \alphab^\Rsm, \qquad \alphab_\Rsm = \alphab^\Lsm.
\ee
Using such notation for the oscillators, we introduce the following ket-vectors:
\beq
\label{27062018-man02-08} && |\phi_s(p,z,\alpha)\rangle = \frac{1}{\sqrt{s!}} \bigl( \alpha_\Lsm^s \phi_s(p,z) +  \alpha_\Rsm^s \phi_{-s}(p,z)\bigr) |0\rangle \,,\qquad s>0\,,
\\
\label{27062018-man02-09} && |\phi_0\rangle = \phi_0(p,z) |0\rangle
\eeq
and in order to treat arbitrary spin fields on an equal footing we use an infinite chain of massless fields which consists of every spin just once. Such chain of massless fields is described by the ket-vector
\be \label{27062018-man02-11}
|\phi(p,z,\alpha)\rangle = \sum_{s=0}^\infty |\phi_s(p,z,\alpha)\rangle\,,
\ee
where the ket-vectors $|\phi_s\rangle$ are defined in \rf{27062018-man02-08},\rf{27062018-man02-09}.
Often, we prefer to use the alternative representation for ket-vector $\phik$ \rf{27062018-man02-11}. Namely, using \rf{27062018-man02-08},\rf{27062018-man02-09}, it is easy to see that  ket-vector $\phik$ \rf{27062018-man02-11} can be represented as
\be \label{27062018-man02-12}
|\phi(p,z,\alpha)\rangle = \sum_{\lambda=-\infty}^\infty \frac{ \alpha_\Hsm^\lambda }{ \sqrt{|\lambda|!} }\phi_\lambda(p,z) |0\rangle \,,\qquad
\ee
where we use a quantity $\alpha_\Hsm^\lambda$ defined by the relations
\be \label{27062018-man02-14}
\alpha_\Hsm^\lambda \equiv \left\{
\begin{array}{l}
\alpha_\Lsm^\lambda  \hspace{1.2cm} \hbox{ for } \lambda > 0\,;
\\[5pt]
1  \hspace{1.5cm} \hbox{ for } \lambda =0\,;
\\[5pt]
\alpha_\Rsm^{-\lambda}  \hspace{1cm} \hbox{ for } \lambda < 0\,.
\end{array}\right.
\ee
It is easy to see that ket-vector \rf{27062018-man02-12} satisfies the algebraic constraint
\be \label{27062018-man02-10}
\alphab^\Rsm\alphab^\Lsm \phik =0\,.
\ee

Throughout the paper we use bra-vector $\langle \phi|$ defined as $\langle \phi(p,z,\alpha)|=(|\phi(p,z,\alpha)\rangle)^\dagger$. Using the expansion \rf{27062018-man02-11}, we get
\be  \label{27062018-man02-14-a1}
\langle\phi(p,z,\alpha)| = \langle 0| \sum_{\lambda=-\infty}^\infty \frac{ \alphab_\Hsm^\lambda }{ \sqrt{|\lambda|!} }\phi_\lambda^\dagger(p,z)   \,,\qquad
\ee
where we use a quantity $\alphab_\Hsm^\lambda$ defined by the relations
\be
\alphab_\Hsm^\lambda \equiv \left\{
\begin{array}{l}
\alphab_\Rsm^\lambda  \hspace{1.2cm} \hbox{ for } \lambda > 0\,;
\\[5pt]
1  \hspace{1.5cm} \hbox{ for } \lambda =0\,;
\\[5pt]
\alphab_\Lsm^{-\lambda}  \hspace{1cm} \hbox{ for } \lambda < 0\,.
\end{array}\right.
\ee

\noindent {\bf Field-theoretical realization of so(3,2) algebra}. Now our aim is to provide a field theoretical realization of the $so(3,2)$ algebra on the space of massless AdS fields. In our approach, massless AdS fields are described by the ket-vector $\phik$ \rf{27062018-man02-12}.  A realization of kinematical generators \rf{25062018-man02-06} and dynamical generators \rf{25062018-man02-07} in terms of differential operators acting on the ket-vector $|\phi\rangle$ is given by
\beq
&& \hspace{-2cm} \hbox{\it Kinematical generators}:
\nonumber\\
\label{27062018-man02-15} && P^1=p^1\,,   \hspace{4cm}   P^+=\beta\,,
\\
&& J^{+1}= {\rm i} x^+ p^1 + \partial_{p^1}\beta\,, \hspace{2cm} J^{+-} = {\rm i}x^+P^- + \partial_\beta \beta\,,
\\
&& D = {\rm i} x^+ P^- -\partial_\beta \beta - \partial_{p^1}p^1 +z\partial_z + 1\,,
\\
\label{27062018-man02-20} && K^+ = \frac{1}{2}(2{\rm i} x^+ \partial_\beta
- \partial_{p^1}^2 +z^2)\beta +  {\rm i} x^+ D\,,
\\
\label{27062018-man02-21} && K^1 = \half (2\irm x^+ \partial_\beta - \partial_{p^1}^2 +z^2)p^1
- \partial_{p^1} D  + M^{z1} z + {\rm i} M^{1-}x^+\,,
\\
&& \hspace{-2cm} \hbox{\it Dynamical generators}:
\nonumber\\
\label{27062018-man02-22} && P^- = - \frac{p^1p^1}{2\beta} + \frac{\partial_z^2}{2\beta}\,,
\\
\label{27062018-man02-23} && J^{-1} = -\partial_\beta p^1 + \partial_{p^1} P^- + M^{-1}  \,,
\\
\label{27062018-man02-24} && K^- = \half (2 \irm  x^+ \partial_\beta - \partial_{p^1}^2 +z^2)P^-
- \partial_\beta D  + \frac{1}{\beta}(\partial_z\partial_{p^1} -zp^1) M^{z1} + \frac{1}{\beta}B\,,
\eeq
where we use the notation
\beq
\label{27062018-man02-25} && M^{-1} \equiv  - M^{z1}\frac{\partial_z}{\beta} \,, \qquad B \equiv - M^{z1} M^{z1}\,,\qquad M^{1-} = - M^{-1}\,,
\\
\label{27062018-man02-26} && \beta\equiv p^+\,,\qquad
\partial_\beta\equiv \partial/\partial \beta\,, \quad
\partial_{p^1}\equiv \partial/\partial p^1\,.
\eeq
In \rf{27062018-man02-25}  and below, a quantity $M^{z1}$ stands for a spin operator of the $so(2)$ algebra. On space of ket-vectors \rf{27062018-man02-11}, the operator $M^{z1}$ is realized as
\be
\label{27062018-man02-28} M^{z1} = M^{\Rsm\Lsm}\,, \qquad
M^{\Rsm\Lsm} \equiv  \alpha^\Rsm \alphab^\Lsm - \alpha^\Lsm \alphab^\Rsm\,.
\ee

Relations \rf{27062018-man02-15}-\rf{27062018-man02-28} provide the realization of the generators of the $so(3,2)$ algebra in terms of differential operators acting on the ket-vector $\phik$ \rf{27062018-man02-12}. Using these relations, we are ready to present a field theoretical realization for the generators of the $so(3,2)$ algebra in terms of the ket-vectors $|\phi\rangle$ \rf{27062018-man02-12}. This is to say that, at the quadratic level, a field theoretical realization of the generators given in \rf{25062018-man02-06},\rf{25062018-man02-07} takes the form
\be \label{27062018-man02-29}
G_\smpt
=\int \beta dz d^2 p\,
\langle\phi(p,z,\alpha)| G |\phi(p,z,\alpha)\rangle\,, \qquad
d^2p \equiv
d\beta dp^1\,,
\ee
where $G_\smpt$ stands for the field theoretical generators, while
$G$ stands for the differential operators presented in \rf{27062018-man02-15}-\rf{27062018-man02-28}.

By definition, the ket-vector $|\phi\rangle$ satisfies the Poisson-Dirac commutation relations
\be \label{27062018-man02-30}
[\,|\phi(p,z,\alpha)\rangle\,,\,|\phi(p^\prime\,,z',\alpha^\prime)\rangle\,]
\bigl|_{{\rm equal}\, x^+}=\bigr.\frac{\delta^2(p+p^\prime)}{2\beta} \delta(z-z')\Pi(\alpha,\alpha')\,,
\ee
where $\Pi(\alpha,\alpha')$ stands for the projector on space of the ket-vector given in \rf{27062018-man02-12}. Using relations \rf{27062018-man02-29} and \rf{27062018-man02-30}, we check the standard commutation relation
\be \label{27062018-man02-31}
[ |\phi\rangle,G_\smpt\,]\bigl|_{{\rm equal}\, x^+}
= G
|\phi\rangle\,.
\ee
In terms of the component fields $\phi_\lambda(p,z)$, the Poisson-Dirac commutation relations take the form
\be  \label{27062018-man02-31-b1}
[\phi_\lambda(p,z),\phi_{\lambda'}(p',z')]\bigl|_{{\rm equal}\, x^+} = \frac{1}{2\beta}\delta^2(p+p')\delta(z-z')\delta_{\lambda+\lambda',0}\,.
\ee
All commutators between fields $\phi_\lambda(p,z)$ and their hermitian conjugated $\phi_\lambda^\dagger(p,z) \equiv (\phi_\lambda(p,z))^\dagger$ are obtained from \rf{27062018-man02-31-b1} by using the hermicity condition $\phi_{-\lambda}(-p,z) = \phi_\lambda^\dagger(p,z)$ \rf{27062018-man02-04}.

For the reader convenience, we note that, in the framework of the Lagrangian approach, the light-cone gauge action takes the form
\be  \label{27062018-man02-31-a1}
S = \int dx^+ dz d^2 p\,\, \langle \phi(p,z,\alpha)|{\rm
i}\, \beta
\partial^- |\phi(p,z,\alpha)\rangle +\int dx^+ P^-\,,
\ee
where $\partial^-\equiv\partial/\partial x^+$ and $P^-$ is the Hamiltonian. Expressions for the
action given in \rf{27062018-man02-31-a1} is valid both for the free and interacting light-cone gauge fields. In the theory of free light-cone gauge fields, the Hamiltonian $P^-$ is obtained by plugging the operator $P^-$ \rf{27062018-man02-22}  into \rf{27062018-man02-29}.
\be  \label{27062018-man02-31-a2}
S_\smpt  = \half \int dx^+ dz d^2 p\,\, \langle \phi(p,z,\alpha)| \bigl(2\irm \beta \partial^-  - p^1p^1 + \partial_z^2\bigr) |\phi(p,z,\alpha)\rangle\,.
\ee

Incorporation of a internal symmetry for massless AdS fields can be done by analogy with the Chan--Paton method used in string theory \cite{Paton:1969je} and in massless arbitrary spin fields in \cite{Metsaev:1991nb} (see remark at the end of Sec.6 in this paper).

%%%%%%%%%%%%%%%%%%%%%%%%%%%%%%%%%%%%%%%%%%%%%%%%%%%%%%%%%%%%%%%%%%%%%%%%%%%%%%%%%%%%%%%%%%%
%%%%%%%%%%%%%%%%%%%%%%%%%%%%%%%%%%%%%%%%%%%%%%%%%%%%%%%%%%%%%%%%%%%%%%%%%%%%%%%%%%%%%%%%%%%
\newsection{ \large Restrictions imposed on $n$-point interaction vertices by kinematical symmetries of $so(3,2)$ algebra } \label{sec-03}
%%%%%%%%%%%%%%%%%%%%%%%%%%%%%%%%%%%%%%%%%%%%%%%%%%%%%%%%%%%%%%%%%%%%%%%%%%%%%%%%%%%%%%%%%%
%%%%%%%%%%%%%%%%%%%%%%%%%%%%%%%%%%%%%%%%%%%%%%%%%%%%%%%%%%%%%%%%%%%%%%%%%%%%%%%%%%%%%%%%%%

Our aim in this Section is to discuss restrictions imposed on the dynamical generators \rf{25062018-man02-07} by the kinematical symmetries. Namely we are going to find equations obtained from commutators between the kinematical generators given in \rf{25062018-man02-06} and dynamical generators given in \rf{25062018-man02-07}.

In theories of interacting fields propagating in AdS space, the dynamical generators receive corrections involving higher
powers of physical fields. Namely, the dynamical generators \rf{25062018-man02-07} can be presented in the following way
\be \label{30062018-man02-01}
G^\dyn = \sum_{n=2}^\infty G_\smpn^\dyn\,,
\ee
where $G_\smpn^\dyn$ \rf{30062018-man02-01} is a functional
that has $n$ powers of physical fields $\phik$. Dynamical generators at quadratic approximation are given by expressions \rf{27062018-man02-22}-\rf{27062018-man02-24} and \rf{27062018-man02-29}.
Now we discuss the general structure of the generators $ G_\smpn^\dyn$ when $n\geq 3$.
We discuss restrictions obtained from commutators between the kinematical generators \rf{25062018-man02-06} and dynamical generators \rf{25062018-man02-07} in turn.
Note that, in what follows, without loss of generality, we study the commutators of the generators of the $so(3,2)$ algebra for $x^+=0$.

\noindent {\bf $P^1$- and $P^+$ symmetries restrictions}. Using the commutators between the dynamical generators \rf{25062018-man02-07} and the kinematical generators $P^1$ and $P^+$, we find that the dynamical generators $G_\smpn^\dyn$ with $n\geq 3$ can be presented as
\beq
\label{30062018-man02-02} && P_\smpn^- = \int\!\! d\Gamma_n \,\, \langle \Phi_\smpn ||p_\smpn^-\rangle_\delta^\vph \,,
\\
\label{30062018-man02-03} && J_\smpn^{-1} = \int \!\! d\Gamma_n\,\, \langle \Phi_\smpn | j_\smpn^{-1}\rangle_\delta^\vph + \bigl( \Xbf^1 \langle \Phi_\smpn | \bigr)|p_\smpn^-\rangle_\delta^\vph \,,
\\
\label{30062018-man02-04} && K_\smpn^- =\int\!\! d\Gamma_n\,\, \langle \Phi_\smpn |k_\smpn^-\rangle_\delta^\vph - \bigl( \Xbf^1 \langle \Phi_\smpn| \bigr) |j_\smpn^{-1}\rangle_\delta^\vph - \half \bigl( \Xbf^1 \Xbf^1 \langle \Phi_\smpn| \bigr)|p_\smpn^-\rangle_\delta^\vph  \,,
\eeq
where bra and ket-vectors appearing in  \rf{30062018-man02-02}-\rf{30062018-man02-04} are defined as
\beq
\label{30062018-man02-05} && \langle \Phi_\smpn| \equiv \prod_{a=1}^n \langle \phi(p_a,z_a,\alpha_a)|\,,\qquad\qquad
\\
&& |p_\smpn^-\rangle_\delta^\vph = \int dz\, | p_\smpn^-\rangle\, \delta_z\,,
\nonumber\\
\label{30062018-man02-06} && \hspace{2.7cm} |p_\smpn^-\rangle =  p_\smpn^- (p_a^1,\partial_{z_a},\beta_a,z, \alpha_a) |0\rangle \,,
\\
&&   |j_\smpn^{-1}\rangle_\delta^\vph = \int dz\, |j_\smpn^{-1} \rangle \, \delta_z\,,
\nonumber\\
\label{30062018-man02-07} && \hspace{2.7cm}   |j_\smpn^{-1}\rangle = j_\smpn^{-1} (p_a^1,\partial_{z_a},\beta_a,z, \alpha_a) |0\rangle \,,
\\
&&   |k_\smpn^-\rangle_\delta^\vph = \int dz\, |k_\smpn^- \rangle\, \delta_z \,,
\nonumber\\
\label{30062018-man02-08} &&  \hspace{2.7cm}  |k_\smpn^-\rangle =  k_\smpn^- (p_a^1,\partial_{z_a},\beta_a,z, \alpha_a) |0\rangle \,,
\eeq
while the remaining quantities appearing in \rf{30062018-man02-02}-\rf{30062018-man02-08} are given by
\beq
\label{30062018-man02-09} && d\Gamma_n \equiv (2\pi)^2  \delta^2(\sum_{a=1}^np_a)
\prod_{a=1}^n \frac{d^2 p_a}{ 2\pi }dz_a \,,\qquad d^2p_a = dp_a^1d\beta_a\,,
\\
\label{30062018-man02-10} && \Xbf^1 \equiv - \frac{1}{n}\sum_{a=1}^n \partial_{p_a^1}\,,\qquad \partial_{p_a^1} \equiv \partial/\partial p_a^1\,,\qquad \partial_{z_a} \equiv \partial/\partial z_a\,,
\\
\label{30062018-man02-11} && \delta_z \equiv  \prod_{a=1}^n \delta(z-z_a)\,, \qquad |0\rangle \equiv \prod_{a=1}^n |0\rangle_a\,.
\eeq

In relations \rf{30062018-man02-05}-\rf{30062018-man02-11} and below, we use the indices $a,b=1,\ldots,n$ to label $n$ interacting
fields. The Dirac $\delta$- functions appearing in \rf{30062018-man02-09}
imply conservation laws for the momenta $p_a^1$ and
$\beta_a$. Densities $p_\smpn^-$, $j_\smpn^{-1}$, and  $k_\smpn^-$ appearing on r.h.s. in
relations \rf{30062018-man02-06}-\rf{30062018-man02-08} depend on the momenta $p_a^1$, $\beta_a$, the radial derivatives $\partial_{z_a}$, the radial coordinate $z$, and spin variables
denoted by $\alpha_a$ in this paper. We note that the shortcut $\alpha_a$ stands for the oscillators $\alpha_a^\Rsm$, $\alpha_a^\Lsm$.
Sometimes, the density $p_\smpn^-$ will be referred to as an $n$-point interaction vertex (or cubic interaction vertex when $n=3$).

\noindent {\bf $J^{+-}$-symmetry restrictions}. Using the commutators between the dynamical generators \rf{25062018-man02-07} and the kinematical generator $J^{+-}$, we get the equations
\beq
\label{30062018-man02-14} && \sum_{a=1}^n \beta_a\partial_{\beta_a} \, |p_\smpn^-\rangle_\delta^\vph = 0 \,,
\\
\label{30062018-man02-15} && \sum_{a=1}^n \beta_a\partial_{\beta_a} \, |j_\smpn^{-1}\rangle_\delta^\vph = 0 \,,
\\
\label{30062018-man02-16} &&  \sum_{a=1}^n \beta_a\partial_{\beta_a} \, |k_\smpn^-\rangle_\delta^\vph = 0 \,.
\eeq

\noindent {\bf $D$-symmetry restrictions}. Using the commutators between the dynamical generators \rf{25062018-man02-07} and the kinematical generator $D$, we get the equations
\beq
\label{30062018-man02-17} && \sum_{a=1}^n (\beta_a\partial_{\beta_a} + p_a^1\partial_{p_a^1} - \partial_{z_a} z_a )|p_\smpn^-\rangle_\delta^\vph = (3-n) |p_\smpn^-\rangle_\delta^\vph\,,
\\
\label{30062018-man02-18} && \sum_{a=1}^n (\beta_a\partial_{\beta_a} + p_a^1\partial_{p_a^1} - \partial_{z_a} z_a ) |j_\smpn^{-1}\rangle_\delta^\vph = (2-n) |j_\smpn^{-1}\rangle_\delta^\vph\,,
\\
\label{30062018-man02-19} && \sum_{a=1}^n (\beta_a\partial_{\beta_a} + p_a^1\partial_{p_a^1} - \partial_{z_a} z_a ) |k_\smpn^-\rangle_\delta^\vph = (1-n) |k_\smpn^-\rangle_\delta^\vph\,.
\eeq

\noindent {\bf $J^{+1}$-symmetry restrictions}. Using the commutators between the dynamical generators \rf{25062018-man02-07} and the kinematical generator $J^{+1}$, we find the equations
\beq
\label{30062018-man02-19-a1} && \sum_{a=1}^n \beta_a\partial_{p_a^1} \, |p_\smpn^-\rangle_\delta^\vph = 0 \,,
\\
\label{30062018-man02-19-a2} && \sum_{a=1}^n \beta_a\partial_{p_a^1} \, |j_\smpn^{-1}\rangle_\delta^\vph = 0 \,,
\\
\label{30062018-man02-19-a3} &&  \sum_{a=1}^n \beta_a\partial_{p_a^1} \, |k_\smpn^-\rangle_\delta^\vph = 0 \,.
\eeq

\noindent {\bf $K^+$-symmetry restrictions}. Using the commutators between the dynamical generators \rf{25062018-man02-07} and the kinematical generator $K^+$, we get the following equations
\beq
\label{30062018-man02-24} && \sum_{a=1}^n \beta_a (z_a^2 - \partial_{p_a^1}\partial_{p_a^1} ) |p_\smpn^-\rangle_\delta^\vph =  0 \,,
\\
\label{30062018-man02-25} && \sum_{a=1}^n \beta_a (z_a^2 - \partial_{p_a^1}\partial_{p_a^1} ) |j_\smpn^{-1}\rangle_\delta^\vph =  0 \,,
\\
\label{30062018-man02-26} && \sum_{a=1}^n \beta_a (z_a^2 - \partial_{p_a^1}\partial_{p_a^1} ) |k_\smpn^-\rangle_\delta^\vph =  0 \,,
\eeq
where for the derivation of equations in \rf{30062018-man02-24}-\rf{30062018-man02-26} we use  equations \rf{30062018-man02-19-a1}-\rf{30062018-man02-19-a3}.

\noindent {\bf $K^1$-symmetry restrictions}. Using the commutators between the dynamical generators \rf{25062018-man02-07} and the kinematical generator $K^1$, we get the equations
\beq
\label{30062018-man02-27} &&     \Kbf^{1\dagger } |p_\smpn^-\rangle_\delta^\vph +  \Xbf^1 |p_\smpn^-\rangle_\delta^\vph  - |j_\smpn^{-1}\rangle_\delta^\vph = 0 \,,
\\
\label{30062018-man02-28} &&  \Kbf^{1\dagger } |j_\smpn^{-1}\rangle_\delta^\vph
- [\Kbf^{1\dagger},\Xbf^1] |p_\smpn^-\rangle_\delta^\vph   + \half \Xbf^1 \Xbf^1 |p_\smpn^-\rangle_\delta^\vph   +  |k_\smpn^-\rangle_\delta^\vph = 0 \,, \qquad
\\
\label{30062018-man02-29} &&  \Kbf^{1\dagger } |k_\smpn^-\rangle_\delta^\vph   - \Xbf^1 |k_\smpn^-\rangle_\delta^\vph +  [\Kbf^{1\dagger},\Xbf^1] |j_\smpn^{-1}\rangle_\delta^\vph   - \half \Xbf^1 \Xbf^1 |j_\smpn^{-1}\rangle_\delta^\vph  = 0\,,
\eeq
where $\Xbf^1$ is defined in \rf{30062018-man02-10}, while $\Kbf^{1\dagger}$ is defined as
\beq
\label{30062018-man02-30} \Kbf^{1\dagger} & \equiv & \sum_{a=1}^n  K_a^{1\dagger}\,,
\\
\label{30062018-man02-31} && K_a^{1\dagger} = \frac{1}{2}p_a^1 (z_a^2 -\partial_{p_a^1}^2) + D_a^\dagger
\partial_{p_a^1} + M_a^{z1} z_a\,,
\\
\label{30062018-man02-32} && D_a^\dagger =  \beta_a \partial_{\beta_a}  +  p_a^1 \partial_{p_a^1} -\partial_{z_a} z_a + 1\,.
\eeq
Using relations given in \rf{30062018-man02-10}, \rf{30062018-man02-30}, we find the helpful relation
\be \label{30062018-man02-33}
[\Kbf^{1\dagger},\Xbf^1]  = \frac{1}{2n}   \sum_{a=1}^n  \bigl( z_a^2  +  \partial_{p_a^1}\partial_{p_a^1} \bigr) \,.
\ee
Note that, for the derivation of equations given in \rf{30062018-man02-27}-\rf{30062018-man02-29}, we use restrictions imposed by $D$-symmetry \rf{30062018-man02-17}-\rf{30062018-man02-19} and restrictions imposed by $J^{+1}$-symmetry \rf{30062018-man02-19-a1}-\rf{30062018-man02-19-a3}.

We summarize our consideration in this section by the following two remarks.

\noindent \ibf)  The use of the commutators of the dynamical generators \rf{25062018-man02-07} with the kinematical generators $P^1$, $P^+$ leads to relations \rf{30062018-man02-02}-\rf{30062018-man02-04}, while the use of the
commutators of the dynamical generators \rf{25062018-man02-07} with the kinematical generators $J^{+-}$, $J^{+1}$, $D$, $K^+$, $K^1$ leads to the equations given in \rf{30062018-man02-14}-\rf{30062018-man02-29}.

\noindent \iibf)  Equations \rf{30062018-man02-19-a1}-\rf{30062018-man02-19-a3} tell us that the densities $p_\smpn^-$, $j_\smpn^{-1}$, and $k_\smpn^-$ depend on the momenta $p_a^1$ through the new momenta $\Po_{ab}^1$ defined by the relation
\be \label{30062018-man02-20}
\Po_{ab}^1 \equiv p_a^1 \beta_b - p_b^1 \beta_a\,.
\ee
In other words, the densities $p_\smpn^-$, $j_\smpn^{-1}$, $k_\smpn^-$ appearing in \rf{30062018-man02-06}-\rf{30062018-man02-08} turn out to be functions of $\Po_{ab}^1$ in place of $p_a^1$,
\beq
\label{30062018-man02-21} && p_\smpn^- = p_\smpn^- (\Po_{ab}^1,\partial_{z_a},\beta_a, z, \alpha_a)\,,
\\
\label{30062018-man02-22}  &&  j_\smpn^{-1} = j_\smpn^{-1} (\Po_{ab}^1,\partial_{z_a},\beta_a, z, \alpha_a)\,,
\\
\label{30062018-man02-23} &&  k_\smpn^- = k_\smpn^- (\Po_{ab}^1,\partial_{z_a},\beta_a,z,\alpha_a)\,.
\eeq

Using the momentum conservation laws, we check that not all momenta $\Po_{ab}^1$ \rf{30062018-man02-20} are independent. Namely, it easy to check that, for the $n$-point
vertex, there are $n-2$ independent momenta $\Po_{ab}^1$. This implies that, for the case of $n=3$, there is only one independent $\Po_{ab}^1$. This considerably simplifies analysis of equations for cubic densities $p_\smp3^-$, $j_\smp3^{-1}$, $k_\smp3^-$.

%%%%%%%%%%%%%%%%%%%%%%%%%%%%%%%%%%%%%%%%%%%%%%%%%%%%%%%%%%%%%%%%%%%%%%%%%%%%%%%%%%%%%%%%%%%
%%%%%%%%%%%%%%%%%%%%%%%%%%%%%%%%%%%%%%%%%%%%%%%%%%%%%%%%%%%%%%%%%%%%%%%%%%%%%%%%%%%%%%%%%%%
\subsection{ \large Restrictions imposed on cubic interaction vertices by kinematical symmetries of $so(3,2)$ algebra } \label{sec-03-01}
%%%%%%%%%%%%%%%%%%%%%%%%%%%%%%%%%%%%%%%%%%%%%%%%%%%%%%%%%%%%%%%%%%%%%%%%%%%%%%%%%%%%%%%%%%
%%%%%%%%%%%%%%%%%%%%%%%%%%%%%%%%%%%%%%%%%%%%%%%%%%%%%%%%%%%%%%%%%%%%%%%%%%%%%%%%%%%%%%%%%%

\noindent {\bf $J^{+1}$-symmetry restrictions}. As we have already said, for cubic vertices, the momenta $\Po_{12}^1$, $\Po_{23}^1$, $\Po_{31}^1$ are not independent. Namely, using the momentum conservation laws
\be \label{30062018-man02-34}
p_1^1 + p_2^1 + p_3^1 = 0\,, \qquad \quad \beta_1 +\beta_2 +\beta_3 =0 \,,
\ee
it is easy to check that the momenta $\Po_{12}^1$, $\Po_{23}^1$, $\Po_{31}^1$ are expressed in terms of a new momentum $\Po^1$ as
\be \label{30062018-man02-35}
\Po_{12}^1 =\Po_{23}^1 = \Po_{31}^1 = \Po^1 \,,
\ee
where the new momentum $\Po^i$ is defined by the relations
\be  \label{30062018-man02-36}
\Po^1 \equiv \frac{1}{3}\sum_{a=1,2,3}\check{\beta}_a p_a^1\,, \qquad
\check{\beta}_a\equiv \beta_{a+1}-\beta_{a+2}\,, \quad \beta_a\equiv
\beta_{a+3}\,.
\ee
The use of the momentum $\Po^1$ \rf{30062018-man02-34} is advantageous because this momentum is manifestly invariant under cyclic permutations of the external line indices
$1,2,3$. Thus, the densities $p_\smp3^-$, $j_\smp3^{-1}$, and $k_\smp3^-$ are eventually functions of $\Po^1$, $\partial_{z_a}$, $\beta_a$, $z$ and $\alpha_a$:
\beq
\label{30062018-man02-37} && p_\smp3^- = p_\smp3^-(\Po^1,\partial_{z_a},\beta_a,z,\alpha_a)\,,
\\
\label{30062018-man02-38} && j_\smp3^{-1} = j_\smp3^{-1}(\Po^1,\partial_{z_a},\beta_a,z,\alpha_a)\,,
\\
\label{30062018-man02-39} && k_\smp3^- = k_\smp3^-(\Po^1,\partial_{z_a},\beta_a,z,\alpha_a)\,.
\eeq

Now our aim is to represent the kinematical symmetry equations
\rf{30062018-man02-14}-\rf{30062018-man02-19} and \rf{30062018-man02-24}-\rf{30062018-man02-29} in terms of the densities given in \rf{30062018-man02-37}-\rf{30062018-man02-39}. To this end we should just plug \rf{30062018-man02-37}-\rf{30062018-man02-39} in \rf{30062018-man02-14}-\rf{30062018-man02-19} and \rf{30062018-man02-24}-\rf{30062018-man02-29} and, upon differentiating over $p_a^1$ and $\beta_a$, we should take into account the dependence of $\Po^1$ on $p_a^1$ and $\beta_a$ \rf{30062018-man02-36}. Doing so, we get kinematical symmetry equations for the densities given in \rf{30062018-man02-37}-\rf{30062018-man02-39}. We now present the equations obtained in turn.

\noindent {\bf $J^{+-}$-symmetry restrictions}. For $n=3$, equations \rf{30062018-man02-14}-\rf{30062018-man02-16} lead to the following equations for densities \rf{30062018-man02-37}-\rf{30062018-man02-39}:
\beq
\label{30062018-man02-40} && \bigl(\Po^1 \partial_{\Po^1} + \sum_{a=1,2,3} \beta_a\partial_{\beta_a} \bigr) \,|p_\smp3^-\rangle_\delta^\vph  = 0\,,
\\
\label{30062018-man02-41} && \bigl( \Po^1 \partial_{\Po^1} + \sum_{a=1,2,3} \beta_a\partial_{\beta_a} \bigr)\,|j_\smp3^{-1}\rangle_\delta^\vph  = 0\,,
\\
\label{30062018-man02-42} && \bigl( \Po^1 \partial_{\Po^1} + \sum_{a=1,2,3} \beta_a\partial_{\beta_a} \bigr)\,|k_\smp3^-\rangle_\delta^\vph  = 0\,.
\eeq

\noindent {\bf $D$-symmetry restrictions}. For $n=3$, equations \rf{30062018-man02-17}-\rf{30062018-man02-19} lead to the following equations for densities \rf{30062018-man02-37}-\rf{30062018-man02-39}:
\beq
\label{30062018-man02-43} && \bigl( \Po^1 \partial_{\Po^1} - \sum_{a=1,2,3} \partial_{z_a}z_a\bigr)|p_\smp3^-\rangle_\delta^\vph  = 0\,,
\\
\label{30062018-man02-44} && \bigl( \Po^1 \partial_{\Po^1} +1 - \sum_{a=1,2,3} \partial_{z_a}z_a\bigr)|j_\smp3^{-1}\rangle_\delta^\vph  = 0\,,
\\
\label{30062018-man02-45} && \bigl( \Po^1 \partial_{\Po^1} +2 - \sum_{a=1,2,3} \partial_{z_a}z_a\bigr)|k_\smp3^-\rangle_\delta^\vph  = 0\,.
\eeq
Note that, for the derivation of equations \rf{30062018-man02-43}-\rf{30062018-man02-45}, we use equations \rf{30062018-man02-40}-\rf{30062018-man02-42}.

\noindent {\bf $K^+$-symmetry restrictions}. For $n=3$, equations \rf{30062018-man02-24}-\rf{30062018-man02-26} lead to the following equations for densities \rf{30062018-man02-37}-\rf{30062018-man02-39}:
\beq
\label{30062018-man02-46} && ( \beta \partial_{\Po^1}\partial_{\Po^1} +
\sum_{a=1,2,3} \beta_a  z_a^2 \bigr) |p_\smp3^-\rangle_\delta^\vph =  0 \,,
\\
\label{30062018-man02-47} && ( \beta \partial_{\Po^1}\partial_{\Po^1} +
\sum_{a=1,2,3} \beta_a  z_a^2 \bigr) |j_\smp3^{-1}\rangle_\delta^\vph =  0 \,,
\\
\label{30062018-man02-48} && ( \beta \partial_{\Po^1}\partial_{\Po^1} +
\sum_{a=1,2,3} \beta_a  z_a^2 \bigr) |k_\smp3^-\rangle_\delta^\vph =  0 \,,
\eeq
where $\beta \equiv \beta_1\beta_2\beta_3$.

\noindent {\bf $K^1$-symmetry restrictions}. For $n=3$, equations \rf{30062018-man02-27}-\rf{30062018-man02-29} lead to the following equations for densities \rf{30062018-man02-37}-\rf{30062018-man02-39}:
\beq
\label{30062018-man02-49} &&  \Kbf^{1\dagger } |p_\smp3^-\rangle_\delta^\vph - |j_\smp3^{-1}\rangle_\delta^\vph    = 0 \,,
\\
\label{30062018-man02-50} &&  \Kbf^{1\dagger } |j_\smp3^{-1}\rangle_\delta^\vph  -  [\Kbf^{1\dagger},\Xbf^1] |p_\smp3^-\rangle_\delta^\vph      +   |k_\smp3^-\rangle_\delta^\vph  = 0 \,, \qquad
\\
\label{30062018-man02-51} &&  \Kbf^{1\dagger } |k_\smp3^-\rangle_\delta^\vph     +  [\Kbf^{1\dagger},\Xbf^1] |j_\smp3^{-1}\rangle_\delta^\vph      = 0\,,
\eeq
where, in \rf{30062018-man02-49}-\rf{30062018-man02-51}, we use the realization of the operators $\Kbf^{1\dagger}$ and $[\Kbf^{1\dagger},\Xbf^1]$  on space of densities \rf{30062018-man02-37}-\rf{30062018-man02-39},
\beq
\label{30062018-man02-52} && \Kbf^{1\dagger}   =    \bigl( \No_\beta -\frac{1}{3}\sum_{a=1,2,3} \check{\beta}_a
z_a\partial_{z_a} \bigr) \partial_{\Po^1} - \frac{\Po^1}{6\beta}\sum_{a=1,2,3}
\beta_a \check\beta_a z_a^2  + \sum_{a=1,2,3} z_a
M_a^{z1}\,,
\\
\label{30062018-man02-53} && [\Kbf^{1\dagger},\Xbf^1] =  \frac{\Delta_\beta}{18}   \partial_{\Po^1}\partial_{\Po^1}  + \frac{1}{6} \sum_{a=1,2,3} z_a^2\,,
\\
\label{30062018-man02-55} && \No_\beta   \equiv \frac{1}{3}\sum_{a=1,2,3} \check\beta_a \beta_a \partial_{\beta_a}\,,
\qquad \Delta_\beta \equiv \sum_{a=1,2,3} \beta_a^2\,, \qquad \beta \equiv \beta_1\beta_2\beta_3\,.
\eeq
For the derivation of the realizations in \rf{30062018-man02-52},\rf{30062018-man02-53}, we use the definitions given in \rf{30062018-man02-30}-\rf{30062018-man02-33} and the relation $[\Xbf^1,\Po^1]=0$.

From \rf{30062018-man02-49},\rf{30062018-man02-50}, we see that the ket-vectors $|j_\smp3^{-1}\rangle_\delta^\vph$ and $|k_\smp3^-\rangle_\delta^\vph$ are entirely expressed in terms of the $|p_\smp3^-\rangle_\delta^\vph$. Using such representation for the $|j_\smp3^{-1}\rangle_\delta^\vph$ and $|k_\smp3^-\rangle_\delta^\vph$ in terms of the $|p_\smp3^-\rangle$, we verify that, if the vertex $|p_\smp3^-\rangle_\delta^\vph$ satisfies the $J^{+-}$-, $D$-, $K^+$-symmetry equations  \rf{30062018-man02-40},\rf{30062018-man02-43},\rf{30062018-man02-46},
then the respective $J^{+-}$-, $D$-, $K^+$-symmetry equations for $|j_\smp3^{-1}\rangle_\delta^\vph$ and $|k_\smp3^-\rangle_\delta^\vph$ in \rf{30062018-man02-40}-\rf{30062018-man02-48} are satisfied automatically. Thus we see that we can restrict ourselves to study the $J^{+-}$-, $D$-, $K^+$-symmetry equations \rf{30062018-man02-40},\rf{30062018-man02-43},\rf{30062018-man02-46} and $K^1$-symmetry equations \rf{30062018-man02-49}-\rf{30062018-man02-51}.

Kinematical restrictions do not exhaust all restrictions imposed by commutators of the $so(3,2)$ algebra. The remaining restrictions imposed by commutators of the $so(3,2)$ are obtained by considering commutators between dynamical generators. Such commutators are studied in the next section.

%%%%%%%%%%%%%%%%%%%%%%%%%%%%%%%%%%%%%%%%%%%%%%%%%%%%%%%%%%%%%%%%%%%%%%%%%%%%%%%%%%%%%%%%%%%
%%%%%%%%%%%%%%%%%%%%%%%%%%%%%%%%%%%%%%%%%%%%%%%%%%%%%%%%%%%%%%%%%%%%%%%%%%%%%%%%%%%%%%%%%%%
\newsection{ \large Restrictions imposed on cubic interaction vertices by dynamical symmetries of $so(3,2)$ algebra }\label{sec-04}
%%%%%%%%%%%%%%%%%%%%%%%%%%%%%%%%%%%%%%%%%%%%%%%%%%%%%%%%%%%%%%%%%%%%%%%%%%%%%%%%%%%%%%%%%%
%%%%%%%%%%%%%%%%%%%%%%%%%%%%%%%%%%%%%%%%%%%%%%%%%%%%%%%%%%%%%%%%%%%%%%%%%%%%%%%%%%%%%%%%%%

Throughout this paper, restrictions obtained from commutators between the dynamical generators of the $so(3,2)$ algebra \rf{25062018-man02-07} are referred to as dynamical symmetry restrictions. The commutators between the dynamical generators of the $so(3,2)$ algebra \rf{25062018-man02-07} are given by
\beq
\label{02072018-man02-01} && [P^-,J^{-1}] = 0 \,,
\\
\label{02072018-man02-02} && [P^-,K^-] = 0 \,, \qquad [J^{-1}, K^-] = 0\,.
\eeq
Therefore our aim in this section is to obtain restrictions on the densities imposed by the commutators \rf{02072018-man02-01},\rf{02072018-man02-02}. We note then the following important feature of the $so(3,2)$ algebra. It turns out that the study of the commutators \rf{02072018-man02-01},\rf{02072018-man02-02} amounts to the study of the commutator  \rf{02072018-man02-01} and the kinematical $K^1$ symmetry equations \rf{30062018-man02-49}-\rf{30062018-man02-51}.
To see this we note that kinematical $K^1$-symmetry equations \rf{30062018-man02-49}-\rf{30062018-man02-51} amount to the following commutators:
\be \label{02072018-man02-03}
[P^-,K^1] = -J^{-1}\,, \qquad [J^{-1},K^1] = K^- \,, \qquad [K^-,K^1] = 0\,.
\ee
Now, by using the Jacobi identities, it is easy to check that, if commutators \rf{02072018-man02-01}  and \rf{02072018-man02-03} are satisfied, then the commutators \rf{02072018-man02-02} are satisfied automatically. Thus, if we respect $K^1$-symmetry equations \rf{30062018-man02-49}-\rf{30062018-man02-51}, then we can restrict ourselves to study the commutator in \rf{02072018-man02-01}. We now study the commutator \rf{02072018-man02-01} in the cubic approximation.

In the cubic approximation, commutator \rf{02072018-man02-01} takes the form
\be  \label{02072018-man02-04}
[P_\smpt^- ,J_\smp3^{-1}] + [P_\smp3^-,J_\smpt^{-1}]=0\,.
\ee
Using commutator \rf{02072018-man02-04} and relations \rf{30062018-man02-02}-\rf{30062018-man02-04} for $n=3$, we get the following equation for the ket-vectors of the densities given in \rf{30062018-man02-06}-\rf{30062018-man02-08} when $n=3$,
\be  \label{02072018-man02-05}
\Pbf^- |j_\smp3^{-1}\rangle_\delta^\vph  + \Jbf^{-1\dagger} |p_\smp3^-\rangle_\delta^\vph  =0\,,
\ee
where we use the notation
\beq
\label{02072018-man02-06} && \hspace{-0.8cm}   \Pbf^- \equiv \sum_{a=1,2,3} P_a^-\,, \hspace{1.4cm} P_a^- \equiv - \frac{p_a^1 p_a^1}{2\beta_a} + \frac{ \partial_{z_a}^2 }{2\beta_a}\,,
\\
\label{02072018-man02-07} && \hspace{-0.8cm} \Jbf^{-1\dagger} \equiv \sum_{a=1,2,3} J_a^{-1\dagger}  \,, \hspace{1cm} J_a^{-1\dagger} \equiv  p_a^1 \partial_{\beta_a} - P_a^-\partial_{p_a^1}   +
M_a^{z1}\frac{\partial_{z_a}}{\beta_a} \,.\qquad
\eeq
Operators $\Pbf^-$ and $\Jbf^{-1\dagger}$ \rf{02072018-man02-06}, \rf{02072018-man02-07} are expressed in terms of the momenta $p_a^1$, $\beta_a$. Using definition of $\Po^1$ \rf{30062018-man02-36} and plugging \rf{30062018-man02-37},\rf{30062018-man02-38} into \rf{02072018-man02-05}, we can express
the operators $\Pbf^-$ and $\Jbf^{-1\dagger}$ in terms of the momenta $\Po^1$ and $\beta_a$,
\beq
\label{02072018-man02-08} \Pbf^- & = & \frac{\Po^1\Po^1}{2\beta} + \sum_{a=1,2,3}\frac{ \partial_{z_a}^2 }{2\beta_a} \,,
\\
\label{02072018-man02-09} \Jbf^{-1\dagger}  & = & - \frac{1}{\beta} \Po^1 \No_\beta -  \sum_{a=1,2,3} \frac{\check\beta_a}{6\beta_a} \partial_{z_a}^2 \partial_{\Po^1} +  \sum_{a=1,2,3} M_a^{z1}\frac{\partial_{z_a}}{\beta_a}\,,
\eeq
where $\No_\beta$ and $\beta$ are defined in \rf{30062018-man02-55}.

Plugging $|j_\smp3^{-1}\rangle_\delta^\vph$ \rf{30062018-man02-49} into \rf{02072018-man02-05}, we get the equation for the cubic vertex $|p_\smp3^-\rangle_\delta^\vph$,
\be  \label{02072018-man02-11}
\bigl( \Jbf^{-1\dagger} + \Pbf^-\Kbf^{1\dagger} \bigr) |p_\smp3^-\rangle_\delta^\vph  = 0 \,.
\ee

Now we are ready to summarize our study of restrictions imposed on the densities  which are obtained from  commutators of the $so(3,2)$ algebra.

{\bf Complete list of equations imposed on densities by kinematical and dynamical symmetries of $so(3,2)$ algebra}.
As we have already said, for studying the kinematical symmetries of the $so(3,2)$ algebra we can restrict ourselves to study the $J^{+-}$-, $D$-, $K^+$-symmetry equations in \rf{30062018-man02-40},\rf{30062018-man02-43},\rf{30062018-man02-46} and $K^1$-symmetry equations \rf{30062018-man02-49}-\rf{30062018-man02-51}, while, for studying the dynamical symmetries of the $so(3,2)$ algebra, we can restrict ourselves to study equations for the vertex \rf{02072018-man02-11}. This implies that the complete list of restrictions imposed on the densities \rf{30062018-man02-37}-\rf{30062018-man02-39} by the symmetries of the $so(3,2)$ algebra takes the form
\beq
&&  \hspace{-4cm} J^{+-}-\hbox{symmetry}
\nonumber\\[-7pt]
\label{03072018-man02-01} && \bigl( \Po^1 \partial_{\Po^1} + \sum_{a=1,2,3} \beta_a\partial_{\beta_a} \bigr) \,|p_\smp3^-\rangle_\delta^\vph   = 0\,;
\\
&&  \hspace{-4cm} D-\hbox{symmetry}
\nonumber\\[-8pt]
\label{03072018-man02-02} && \bigl( \Po^1 \partial_{\Po^1} - \sum_{a=1,2,3} \partial_{z_a}z_a\bigr)|p_\smp3^-\rangle_\delta^\vph   = 0\,;
\\
&&  \hspace{-4cm} K^+-\hbox{symmetry}
\nonumber\\[-8pt]
\label{03072018-man02-03} && \bigl( \beta \partial_{\Po^1}\partial_{\Po^1} +
\sum_{a=1,2,3} \beta_a  z_a^2 \bigr) |p_\smp3^-\rangle_\delta^\vph  =  0 \,;
\\
&&  \hspace{-4cm} K^1-\hbox{symmetry}
\nonumber\\[-8pt]
\label{03072018-man02-05} && \Kbf^{1\dagger } |p_\smp3^-\rangle_\delta^\vph  - |j_\smp3^{-1}\rangle_\delta^\vph  = 0 \,,
\\
\label{03072018-man02-06} &&  \Kbf^{1\dagger } |j_\smp3^{-1}\rangle_\delta^\vph  -  [\Kbf^{1\dagger},\Xbf^1] |p_\smp3^-\rangle_\delta^\vph      +   |k_\smp3^-\rangle_\delta^\vph  = 0 \,, \qquad
\\
\label{03072018-man02-07} &&  \Kbf^{1\dagger } |k_\smp3^-\rangle_\delta^\vph      +  [\Kbf^{1\dagger},\Xbf^1] |j_\smp3^{-1}\rangle_\delta^\vph      = 0\,;
\\
&&  \hspace{-4cm} P^-,J^{-1}-\hbox{symmetries}
\nonumber\\[-8pt]
\label{03072018-man02-08} && \bigl( \Jbf^{-1\dagger} + \Pbf^-\Kbf^{1\dagger}\bigr) |p_\smp3^-\rangle_\delta^\vph  = 0 \,;
\eeq
where the operators $\Kbf^{1\dagger}$, $[\Kbf^{1\dagger},\Xbf^1]$, $\Pbf^-$, and $\Jbf^{-1\dagger}$ appearing in \rf{03072018-man02-05}-\rf{03072018-man02-08} are given in \rf{30062018-man02-52}, \rf{30062018-man02-53}, \rf{02072018-man02-08} and \rf{02072018-man02-09} respectively.

Equations \rf{03072018-man02-01}-\rf{03072018-man02-08} do not admit to determine the densities  $p_\smp3^-$, $j_\smp3^{-1}$, and $k_\smp3^-$ uniquely. To determine  the densities $p_\smp3^-$,  $j_\smp3^{-1}$, and $k_\smp3^-$ uniquely we should impose some additional restrictions on the cubic vertex $p_\smp3^-$. We now formulate these additional restrictions.

\noindent \ibf) The vertex $p_\smp3^-$ should be finite-order polynomial in the momentum $\Po^1$ and the derivatives $\partial_{z_a}$.

\noindent \iibf) The vertex  $p_\smp3^-$ should satisfy the restriction
\be \label{03072018-man02-09}
|p_\smp3^-\rangle_\delta^\vph  \ne \Pbf^- |V\rangle_\delta^\vph \,, \quad |V\rangle_\delta^\vph  \ \hbox{ is finite-order polynomial in } \Po^1 \hbox{ and } \partial_{z_a}\,,
\ee
where $\Pbf^-$ is given in \rf{02072018-man02-08}.

In the framework of light-cone approach, the assumption \ibf) is a counterpart of locality condition commonly used in a Lorentz covariant approach. We note also that the assumption \iibf) is related to field redefinitions. Namely, if we ignore requirement \rf{03072018-man02-09}, then we get  vertices which can be removed by field redefinitions. As we are interested to deal with the cubic interaction vertices that cannot be removed by field redefinitions, we respect the requirement in \rf{03072018-man02-09}.

To summarize the discussion in this section, we note that, for densities $p_\smp3^-$, $j_\smp3^{-1}$, $k_\smp3^-$ in \rf{30062018-man02-37}-\rf{30062018-man02-39}, equations \rf{03072018-man02-01}-\rf{03072018-man02-08} supplemented by requirements \ibf) and \iibf) constitute the complete system of equations which admit to determine the densities $p_\smp3^-$, $j_\smp3^{-1}$, $k_\smp3^-$ \rf{30062018-man02-37}-\rf{30062018-man02-39} uniquely.

%%%%%%%%%%%%%%%%%%%%%%%%%%%%%%%%%%%%%%%%%%%%%%%%%%%%%%%%%%%%%%%%%%%%%%%%%%%%%%%%%%%%%%%%%%%
%%%%%%%%%%%%%%%%%%%%%%%%%%%%%%%%%%%%%%%%%%%%%%%%%%%%%%%%%%%%%%%%%%%%%%%%%%%%%%%%%%%%%%%%%%%
\newsection{ \large Method for solving equations for cubic interaction vertex } \label{sec-05}
%%%%%%%%%%%%%%%%%%%%%%%%%%%%%%%%%%%%%%%%%%%%%%%%%%%%%%%%%%%%%%%%%%%%%%%%%%%%%%%%%%%%%%%%%%
%%%%%%%%%%%%%%%%%%%%%%%%%%%%%%%%%%%%%%%%%%%%%%%%%%%%%%%%%%%%%%%%%%%%%%%%%%%%%%%%%%%%%%%%%%

Finding solution to the complete system of equations  \rf{03072018-man02-01}-\rf{03072018-man02-08} turns out to be complicated problem. The most difficult point in the analysis of equations \rf{03072018-man02-01}-\rf{03072018-man02-08} is related with the treatment of the radial derivatives $\partial_{z_a}$. We now describe our procedure for solving equations \rf{03072018-man02-01}-\rf{03072018-man02-08}. Our procedure is realized in the following eight steps.

\noindent {\bf Step 1}. Our aim at this step is to introduce a convenient basis for the radial derivatives $\partial_{z_a}$. To this end we introduce the following decomposition of the radial derivatives $\partial_{z_a}$, $a=1,2,3$:
\beq
\label{03072018-man02-14} && \hspace{-1cm} \partial_{z_a} =  (\frac{1}{3} + \frac{\beta_a\Delta_\beta}{18\beta})\Pbf_z
-\frac{\beta_a\check{\beta}_a}{3\beta}\Po_z +\frac{1}{3}\beta_a \PP_z\,,
\\
\label{03072018-man02-15} && \Pbf_z \equiv \sum_{a=1,2,3}\partial_{z_a}\,,
\\
\label{03072018-man02-16} && \Po_z \equiv \frac{1}{3} \sum_{a=1,2,3}\betach_a \partial_{z_a}\,,
\\
\label{03072018-man02-17} && \PP_z \equiv \sum_{a=1,2,3} \frac{\partial_{z_a}}{\beta_a}\,,
\eeq
where $\betach_a,\beta,\Delta_\beta$ are given in \rf{16072018-man02-10}-\rf{16072018-man02-14} in  Appendix  A.
Quantities $\Pbf_z$, $\Po_z$, and $\PP_z$ defined in \rf{03072018-man02-15}-\rf{03072018-man02-17} will be referred to as radial momenta.
Relations \rf{03072018-man02-14}-\rf{03072018-man02-17} describe the one-to-one mapping between the three radial derivatives $\partial_{z_a}$, $a=1,2,3$, and three radial momenta $\Pbf_z$, $\Po_z$, $\PP_z$. Comparing \rf{03072018-man02-16} with \rf{30062018-man02-36}, we see that the radial momentum $\Po_z$ is defined by analogy to the momentum $\Po^1$ \rf{30062018-man02-36}. We will show  that, in view of various reasons, the radial momenta $\Pbf_z$ and $\PP_z$ can be eliminated from our consideration. First, we consider the radial momentum $\Pbf_z$.

Using decomposition \rf{03072018-man02-14} in \rf{30062018-man02-37}, we see that the vertex $p_\smp3^-$ is represented as
\be \label{03072018-man02-18}
p_\smp3^- = p_\smp3^-(\Po^1,\Pbf_z,\Po_z,\PP_z,\beta_a,z,\alpha_a)\,.
\ee
As the vertex $p_\smp3^-$ is a finite-order polynomial in the radial momentum $\Pbf_z$, we can represent the vertex $p_\smp3^-$ as
\be \label{03072018-man02-19}
p_\smp3^- = \sum_{n=0}^N \Pbf_z^n V_n(\Po^1,\Po_z,\PP_z,\beta_a,z,\alpha_a)\,.
\ee
Now taking into account definition of the $\delta_z$ \rf{30062018-man02-11}, we get the relation
\be \label{03072018-man02-20}
\Pbf_z \delta_z = -\partial_z \delta_z\,, \qquad \partial_z = \partial/\partial z\,.
\ee
Using relation \rf{03072018-man02-20} in  \rf{30062018-man02-06}, we see that, up to total derivative, the vertex $|p_\smp3^-\rangle_\delta$ \rf{30062018-man02-06} having density  $p_\smp3^-$ as in \rf{03072018-man02-19} amounts to the vertex $|p_\smp3^-\rangle_\delta$ having the density given by
\be \label{03072018-man02-21}
p_\smp3^- = \sum_{n=0}^N \partial_z^n V_n(\Po^1,\Po_z,\PP_z,\beta_a,z,\alpha_a)\,.
\ee
Relation \rf{03072018-man02-21} implies that the radial momentum $\Pbf_z$ \rf{03072018-man02-14} can be eliminated from our consideration. In other words, without loss of generality, we can restrict our attention to the  vertex which does not depend on the $\Pbf_z$,
\be \label{03072018-man02-22}
p_\smp3^- = p_\smp3^-(\Po^1,\Po_z,\PP_z,\beta_a,z,\alpha_a)\,.
\ee
Using density $p_\smp3$ \rf{03072018-man02-22}, we now proceed to the next step of our procedure.

\noindent {\bf Step 2}. Our aim at this step is to provide solution to the requirement in \rf{03072018-man02-09}. As in the flat space, this requirement can be solved by using field redefinitions. In the framework of light-cone gauge formulation of field dynamics in flat space, the detailed discussion of field redefinitions may be found in Appendix B in Ref.\cite{Metsaev:2005ar}. Analysis of field redefinitions in AdS space follows the pattern of the analysis described in Appendix B in Ref.\cite{Metsaev:2005ar}. Therefore, to avoid the repetitions, we briefly describe result of the analysis.

Under field redefinitions the vertex $p_\smp3^-$ transforms as
\beq
\label{04072018-man02-01} && p_\smp3\rightarrow  p_\smp3^-  - \Pbf^- f\,, \qquad
\\
\label{04072018-man02-02} && f = f(\Po^1,\Po_z,\PP_z,\beta_a,z,\alpha_a)\,,
\eeq
where vertex $f$ given in \rf{04072018-man02-02} describes generating function of field redefinitions (see relations B3 and B15 in Appendix B in Ref.\cite{Metsaev:2005ar}). As seen from \rf{04072018-man02-02}, the vertex $f$ depends on the same variables as the cubic interaction vertex $p_\smp3^-$ in \rf{03072018-man02-21}. Operator  $\Pbf^-$ appearing in \rf{04072018-man02-01} is defined in \rf{02072018-man02-08}. It is easy to check that, on space of the vertex $f$, operator $\Pbf^-$ \rf{02072018-man02-08} is realized as
\be \label{04072018-man02-03}
\Pbf^-  =  \frac{\Po^1\Po^1-\Po_z\Po_z}{2\beta} + \frac{\Delta_\beta}{36\beta}
\partial_z^2 + \frac{1}{3}\PP_z \partial_z\,.
\ee
We now introduce a definition of harmonic vertex. By definition, vertex $p_\smp3^-$ that satisfies the equation
\be \label{04072018-man02-04}
(\partial_{\Po^1}^2 - \partial_{\Po_z}^2) p_\smp3^- = 0
\ee
is refereed to as harmonic vertex. We recall that, by definition, the vertex
$p_\smp3^-$ is a polynomial in the momenta  $\Po^1$, $\Po_z$. As is
well known an arbitrary polynomial in two variables $\Po^1$, $\Po_z$ can be made
a harmonic polynomial in $\Po^1$, $\Po_z$ by adding a suitable
polynomial proportional to $\Po^1\Po^1-\Po_z\Po_z$. From \rf{04072018-man02-01},\rf{04072018-man02-03}, we see that
it is the polynomial proportional to $\Po^1\Po^1-\Po_z\Po_z$ that is generated by field
redefinitions. This implies that, by using field redefinitions \rf{04072018-man02-01}, the vertex
$p_\smp3^-$ can be made to satisfy the harmonic equation \rf{04072018-man02-04}.

Summarizing the two steps above discussed, we note that we are left with vertex \rf{03072018-man02-21} which satisfies the equation \rf{04072018-man02-04}. Such harmonic vertex obviously satisfies the requirement \rf{03072018-man02-09}. Using \rf{03072018-man02-22},\rf{04072018-man02-04}, we now proceed to the next step of our procedure.

\noindent {\bf Step 3}. We now study restrictions imposed on the vertex \rf{03072018-man02-22}
by $K^+$ symmetry equation \rf{03072018-man02-03}. To this end we note that, for the ket-vector $|p_\smp3^-\rangle_\delta^\vph$ \rf{30062018-man02-06} with $p_\smp3^-$ as in \rf{03072018-man02-22}, the following relation holds true
\be \label{04072018-man02-05}
\sum_{a=1,2,3} \beta_a z_a^2 |p_\smp3^-\rangle_\delta^\vph =    \int dz \Bigl( - \beta\partial_{\Po_z}^2 - 6z \partial_{\PP_z}^\vph -\frac{\Delta_\beta}{2\beta}\partial_{\PP_z}^2 \Bigr)p_\smp3^- \delta_z|0\rangle \,.
\ee
Making use of \rf{04072018-man02-04} and \rf{04072018-man02-05}, we note that $K^+$-symmetry equation \rf{03072018-man02-03} amounts to the following equation for $p_\smp3^-$:
\be \label{04072018-man02-06}
\bigl( 6z  + \frac{\Delta_\beta}{2\beta}\partial_{\PP_z}^\vph \bigr)\partial_{\PP_z}^\vph p_\smp3^- = 0  \,.
\ee
For the vertex $p_\smp3^-$,  which by definition is finite-order polynomial in the radial momentum $\PP_z$, equation \rf{04072018-man02-06} implies that $p_\smp3^-$ is independent of the radial momentum $\PP_z$,
\be \label{04072018-man02-07}
p_\smp3^- = p_\smp3^-(\Po^1,\Po_z,\beta_a,z,\alpha_a)\,.
\ee

Thus, at this step of our procedure, we obtain vertex \rf{04072018-man02-07} which is independent of the radial momentum $\PP_z$. Summarizing the three steps of our procedure above discussed  we note that we are left with vertex \rf{04072018-man02-07} which is harmonic with respect to the momenta $\Po^1$, $\Po_z$ \rf{04072018-man02-04}. Using such vertex, we now proceed to the next step of our procedure.

\noindent {\bf Step 4}. Our aim at this step, is to represent $J^{+-}$- and $D$-symmetry equations \rf{03072018-man02-01},\rf{03072018-man02-02} in terms of the harmonic vertex \rf{04072018-man02-07}. Taking into account the definition of $\Po_z$ \rf{03072018-man02-16}, we verify that in terms of the harmonic vertex $p_\smp3^-$ \rf{04072018-man02-07}, equations \rf{03072018-man02-01},\rf{03072018-man02-02}  can be represented as
\beq
\label{04072018-man02-08} && \bigl( N_{\Po^1} + N_{\Po_z} + \sum_{a=1,2,3} \beta_a\partial_{\beta_a} \bigr) \, p_\smp3^- = 0\,,
\\
\label{04072018-man02-09} && \bigl( N_z  - N_{\Po^1} - N_{\Po_z} + 1 \bigr) p_\smp3^- = 0 \,,
\\
\label{04072018-man02-10} && N_z\equiv z\partial_z \,, \qquad N_{\Po^1} \equiv \Po^1 \partial_{\Po^1}\,, \qquad N_{\Po_z} = \Po_z \partial_{\Po_z}^\vph\,.
\eeq

\noindent {\bf Step 5}. At this step we consider equation \rf{03072018-man02-08} for the ket-vector $|p_\smp3^-\rangle_\delta^\vph$ which involves delta-functions $\delta_z$ \rf{30062018-man02-06}. Our aim is to represent equation \rf{03072018-man02-08} in terms of the ket-vector $|p_\smp3^-\rangle$ that does not involve delta functions $\delta_z$ \rf{30062018-man02-06}, where the vertex $p_\smp3^-$ takes the form as in \rf{04072018-man02-07} and satisfies harmonic constraint \rf{04072018-man02-04}. To this end we start with the presenting realization of the operators $\Kbf^{1\dagger}$ \rf{30062018-man02-52}, $\Pbf^-$ \rf{02072018-man02-08},  $\Jbf^{-1\dagger}$ \rf{02072018-man02-09} on space of the harmonic ket-vector  $|p_\smp3^-\rangle$ \rf{30062018-man02-06}, \rf{04072018-man02-04}, \rf{04072018-man02-07},
\beq
\label{04072018-man02-11} \Kbf^{1\dagger} & = & \No_\beta \partial_{\Po^1}  + \frac{\Delta_\beta}{9}\partial_z \partial_{\Po_z} \partial_{\Po^1}   -  \Mo^{z1}\partial_{\Po_z}  + z \Jbf^{z1}\,,
\\
\label{04072018-man02-12} \Pbf^-  & = &   \frac{\Po^1\Po^1-\Po_z\Po_z}{2\beta} + \frac{\Delta_\beta}{36\beta}
\partial_z^2 + \frac{1}{3}\PP_z \partial_z\,,
\\
\label{04072018-man02-14} \Jbf^{-1\dagger}  & = & - \frac{1}{\beta}\Po^1 \No_\beta - \frac{1}{\beta}\Mo^{z1}\Po_z + \frac{1}{3} \MM^{z1} \partial_z +  \frac{\Delta_\beta}{9\beta} \Po_z\partial_{\Po^1}\partial_z +  \frac{\Delta_\beta}{18\beta}  \Jbf^{z1} \partial_z
\nonumber\\
& + &   \frac{\betach}{54\beta} \partial_{\Po^1} \partial_z^2 + \frac{1}{3}\PP_z \Jbf^{z1}\,,
\eeq
where  $\Delta_\beta,\beta,\betach,\No_\beta$ are defined in  \rf{16072018-man02-10}-\rf{16072018-man02-15} in Appendix A and we use the notation
\beq
\label{04072018-man02-15} && \hspace{-1.5cm} \Jbf^{z1} =  - \Po^1 \partial_{\Po_z}^\vph - \Po_z \partial_{\Po^1}  + \Mbf^{z1}\,,
\\
\label{04072018-man02-16} && \hspace{-1.5cm} \Mbf^{z1} = \sum_{a=1,2,3} M_a^{z1}\,, \qquad \Mo^{z1} = \frac{1}{3}\sum_{a=1,2,3} \betach_a M_a^{z1}\,, \qquad \MM^{z1} =  \sum_{a=1,2,3} \frac{1}{\beta_a}M_a^{z1}\,.
\eeq
Using \rf{04072018-man02-11}-\rf{04072018-man02-14}, one can demonstrate that equation \rf{03072018-man02-08} amounts to the following two equations (for details, see Appendix B)
\beq
\label{04072018-man02-19} && \hspace{-1cm} \Bigl( (N_z + 2)\Jbf^{z1}  +  \No_\beta \partial_{\Po^1}\partial_z - \Mo^{z1}\partial_{\Po_z}\partial_z  + \frac{\Delta_\beta}{9}  \partial_{\Po_z} \partial_{\Po^1} \partial_z^2 \Bigr)|p_\smp3^-\rangle = 0 \,,
\\
\label{04072018-man02-20} && \hspace{-1cm} \Bigl( -  \Po_\Thsm^1 \No_\beta -  \Mo^{z1} \Po_{\Thsm,z} + \frac{\beta}{3} \MM^{z1}\partial_z   +  \frac{\Delta_\beta}{9} \Po_{\Thsm,z}\partial_{\Po^1}\partial_z +  \frac{\Delta_\beta}{36}  \Jbf^{z1} \partial_z + \frac{\betach}{54} \partial_{\Po^1}\partial_z^2 \Bigr) |p_\smp3^-\rangle = 0 \,,\qquad
\eeq
where operators $ \Po_\Thsm^1$, $\Po_{\Thsm,z}$ appearing in \rf{04072018-man02-20} are defined as
\beq
\label{04072018-man02-21} && \Po_\Thsm^1 \equiv  \Po^1 - (\Po^1\Po^1 - \Po_z\Po_z) \frac{1}{2N_{\Po^1} + 2N_{\Po_z} +2}  \partial_{\Po^1}\,,
\\
\label{04072018-man02-22} && \Po_{\Thsm,z} \equiv \Po_z + (\Po^1\Po^1 - \Po_z\Po_z) \frac{1}{2N_{\Po^1} + 2N_{\Po_z} +2}  \partial_{\Po_z}^\vph\,,
\eeq
while $N_{\Po^1}$, $N_{\Po_z}$ are defined in \rf{04072018-man02-10}.
For harmonic vertex \rf{04072018-man02-04}, we note the following important relations for the operators $ \Po_\Thsm^1$, $\Po_{\Thsm,z}$,
\be \label{04072018-man02-30}
(\partial_{\Po^1}^2 - \partial_{\Po_z}^2) \Po_\Thsm^1 p_\smp3^- = 0\,, \qquad
(\partial_{\Po^1}^2 - \partial_{\Po_z}^2) \Po_{\Thsm,z} p_\smp3^- = 0\,.
\ee
Relations \rf{04072018-man02-30} tell us that the operators $ \Po_\Thsm^1$, $\Po_{\Thsm,z}$ respect the harmonic condition \rf{04072018-man02-04}.

\noindent {\bf Step 6}. At this step we analyse $K^1$-symmetry equations \rf{03072018-man02-05}-\rf{03072018-man02-07} for ket-vectors $|p_\smp3^-\rangle_\delta^\vph$, $|j_\smp3^{-1}\rangle_\delta^\vph$, $|k_\smp3^-\rangle_\delta^\vph$, which involve delta functions $\delta_z$ \rf{30062018-man02-06}-\rf{30062018-man02-08}. In terms of the ket-vectors $|p_\smp3^-\rangle$, $|j_\smp3^{-1}\rangle$, $|k_\smp3^-\rangle$, which do not involve the delta-functions $\delta_z$ \rf{30062018-man02-06}-\rf{30062018-man02-08}, these equations take the form
\beq
\label{06072018-man02-01} && \Kbf^{1\dagger } |p_\smp3^-\rangle   - |j_\smp3^{-1}\rangle = 0 \,,
\\
\label{06072018-man02-02} && \Kbf^{1\dagger } |j_\smp3^{-1}\rangle       -  \half z^2 |p_\smp3^-\rangle   - \frac{\Delta_\beta}{9}\partial_{\Po^1}^2 |p_\smp3^-\rangle     + |k_\smp3^-\rangle  = 0 \,,
\\
\label{06072018-man02-03} && \Kbf^{1\dagger } |k_\smp3^-\rangle   +  \half z^2 |j_\smp3^{-1}\rangle + \frac{\Delta_\beta}{9}\partial_{\Po^1}^2 |j_\smp3^{-1}\rangle  = 0\,,
\eeq
where realization of the operator $\Kbf^{1\dagger}$ on space of the ket-vectors $|p_\smp3^-\rangle$, $|j_\smp3^{-1}\rangle$, $|k_\smp3^-\rangle$ is given in \rf{04072018-man02-11}, while realization of the operator $[\Kbf^{1\dagger},\Xbf^1]$ is given by
\be \label{06072018-man02-04}
[\Kbf^{1\dagger},\Xbf^1]  = \half z^2   +  \frac{\Delta_\beta}{18} (\partial_{\Po_z}^2 + \partial_{\Po^1}^2)\,.
\ee
Note that, from \rf{06072018-man02-01},\rf{06072018-man02-02}, we learn that $j_\smp3^{-1}$, $k_\smp3^-$ are harmonic functions in $\Po^1$, $\Po_z$ \rf{04072018-man02-04}.

Equations \rf{06072018-man02-01},\rf{06072018-man02-02} are algebraic relations which tell us that ket-vectors  $|j_\smp3^{-1}\rangle$ and $|k_\smp3^-\rangle$ are entirely expressed in terms of the cubic vertex $|p_\smp3^-\rangle$. We note then that, by using equations \rf{04072018-man02-19}, we can  express the operators $\Jbf^{z1}$ as in \rf{04072018-man02-27} in Appendix B. This is to say that, using \rf{04072018-man02-11} and \rf{04072018-man02-27}, we can represent relations \rf{06072018-man02-01},\rf{06072018-man02-02} in a more convenient-to-use form
\beq
\label{06072018-man02-05} && |j_\smp3^{-1}\rangle = \Kbf^{1\dagger } |p_\smp3^-\rangle\,,
\\
\label{06072018-man02-06} && |k_\smp3^-\rangle = \half z^2 |p_\smp3^-\rangle - z \Kbf^{z\dagger } |p_\smp3^-\rangle \,,
\\
\label{06072018-man02-07} &&\hspace{1cm} \Kbf^{1\dagger }  =  \frac{1}{N_z+1} \Bigl( \No_\beta \partial_{\Po^1}  - \Mo^{z1}\partial_{\Po_z}   + \frac{\Delta_\beta}{9}  \partial_{\Po_z} \partial_{\Po^1} \partial_z \Bigr),
\\
\label{06072018-man02-08} &&\hspace{1cm} \Kbf^{z\dagger }=  \frac{1}{N_z+1} \Bigl( \No_\beta \partial_{\Po_z}  - \Mo^{z1}\partial_{\Po^1}   + \frac{\Delta_\beta}{9}  \partial_{\Po_z}^2 \partial_z \Bigr).
\eeq
Using relations \rf{06072018-man02-05},\rf{06072018-man02-06}, we then verify that equation \rf{06072018-man02-03} is satisfied automatically. Thus relations \rf{06072018-man02-05}-\rf{06072018-man02-08} provide the solution to the $K^1$-symmetry equations.

Thus, we reduced the problem of solving of equations for ket-vectors $|p_\smp3^-\rangle_\delta^\vph$, $|j_\smp3^{-1}\rangle_\delta^\vph$, $|k_\smp3^-\rangle_\delta^\vph$ \rf{03072018-man02-01}-\rf{03072018-man02-08} to the problem of solving equations for the harmonic vertex $|p_\smp3^-\rangle$, where the $p_\smp3$ takes the form given in \rf{04072018-man02-07}. The remaining equations for the harmonic vertex $|p_\smp3^-\rangle$ to be studied are given in  \rf{04072018-man02-08},\rf{04072018-man02-09} and \rf{04072018-man02-19},\rf{04072018-man02-20}. Also we expressed the harmonic ket-vectors $|j_\smp3^{-1}\rangle$, $|k_\smp3^-\rangle$ in terms of the harmonic vertex $|p_\smp3^-\rangle$ \rf{06072018-man02-05}-\rf{06072018-man02-08}.

\noindent {\bf Step 7}. Our aim at this step is to represent the remaining equations for the harmonic vertex $|p_\smp3^-\rangle$ given in \rf{04072018-man02-08},\rf{04072018-man02-09} and \rf{04072018-man02-19},\rf{04072018-man02-20} in terms of holomorphic momenta and anti-holomorphic momenta which are defined below. The use of such momenta allows us to introduce holomorphic and anti-holomorphic vertices. It turns out that the remaining equations \rf{04072018-man02-08},\rf{04072018-man02-09} and \rf{04072018-man02-19},\rf{04072018-man02-20} lead to decoupled equations for holomorphic and anti-holomorphic vertices and this simplifies our procedure for finding the cubic vertices.

Holomorphic and anti-holomorphic momenta denoted by $\Po^\Lsm$ and $\Po^\Rsm$ are defined by the relations
\be \label{07072018-man02-01}
\Po^\Lsm = \frac{1}{\sqrt{2}}(\Po^1 + \Po_z)\,,  \qquad  \Po^\Rsm = \frac{1}{\sqrt{2}}(\Po^1 - \Po_z)\,.
\ee
In terms of momenta \rf{07072018-man02-01}, equation \rf{04072018-man02-04} takes the form
\be  \label{07072018-man02-02}
\partial_{\Po^\Lsm}^\vph \partial_{\Po^\Rsm}^\vph p_\smp3^- = 0\,.
\ee
General solution to equation \rf{07072018-man02-02} can be presented as
\beq
\label{07072018-man02-03}  && \hspace{-1.5cm} p_\smp3^- = V_\scalarrm^{000} + V + \Vb\,,
\\
\label{07072018-man02-04} && \hspace{-1.5cm} V_\scalarrm^{000} =  V_\scalarrm^{000}(\beta_a,z,\alpha_a)\,, \qquad V =  V(\Po^\Lsm,\beta_a,z,\alpha_a)\,, \qquad  \Vb = \Vb(\Po^\Rsm,\beta_a,z,\alpha_a)\,,\qquad
\eeq
where vertices $V$ and $\Vb$ do not involve the respective terms of zero-order in $\Po^\Lsm$ and $\Po^\Rsm$. Obviously, for vertices given in \rf{07072018-man02-04}, the $J^{+-}$- and $D$-symmetry equations \rf{04072018-man02-08},\rf{04072018-man02-09} turn out to be decoupled and take the following form:
\beq
\label{14072018-man02-01} && \sum_{a=1,2,3} \beta_a\partial_{\beta_a} V_\scalarrm^{000} = 0\,, \hspace{2.7cm} \bigl( N_z + 1 \bigr) V_\scalarrm^{000} = 0 \,,
\\
\label{14072018-man02-02} && \bigl( N_{\Po^\Lsm}  + \sum_{a=1,2,3} \beta_a\partial_{\beta_a} \bigr) \, V = 0\,, \hspace{1.5cm} \bigl( N_z  - N_{\Po^\Lsm} + 1 \bigr) V = 0 \,,
\\
\label{14072018-man02-03} && \bigl( N_{\Po^\Rsm}  + \sum_{a=1,2,3} \beta_a\partial_{\beta_a} \bigr) \, \Vb = 0\,, \hspace{1.5cm} \bigl( N_z  - N_{\Po^\Rsm} + 1 \bigr) \Vb = 0 \,,
\\
&& \hspace{1cm}  N_z\equiv z\partial_z \,,\qquad N_{\Po^\Lsm} \equiv \Po^\Lsm \partial_{\Po^\Lsm}\,, \qquad N_{\Po^\Rsm} = \Po^\Rsm \partial_{\Po^\Rsm}\,.
\eeq

In Appendix C, we outline the proof of the following two Statements.

\noindent \ibf) Equations \rf{04072018-man02-19},\rf{04072018-man02-20} and \rf{14072018-man02-01}-\rf{14072018-man02-03} lead to the following solution for the vertex $V_\scalarrm^{000}$:
\be \label{15072018-man02-01}
V_\scalarrm^{000} = \frac{C_\scalarrm^{000}}{z}\,,
\ee
where $C_\scalarrm^{000}$ is a constant parameter (a coupling constant of three scalar fields).

\noindent \iibf) Equation \rf{04072018-man02-19} leads to the following decoupled equations for holomorphic and anti-holomorphic ket-vectors $|V\rangle = V|0\rangle$, $|\Vb\rangle = \Vb|0\rangle$:
\beq
\label{07072018-man02-05}  && \Bigl( \Jbf^{\Rsm\Lsm} + \frac{1}{N_z+2}\Bigl( \frac{1}{\sqrt{2}} \No_\beta \partial_{\Po^\Lsm} \partial_z  - \frac{1}{ \sqrt{2} }\Mo^{\Rsm\Lsm}\partial_{\Po^\Lsm} \partial_z + \frac{\Delta_\beta}{18}   \partial_{\Po^\Lsm}^2    \partial^2_z\Bigr) \Bigr) |V\rangle   = 0\,,
\\
\label{07072018-man02-06} && \Bigl(\Jbf^{\Rsm\Lsm} + \frac{1}{N_z+2}\Bigl( \frac{1}{\sqrt{2}} \No_\beta \partial_{\Po^\Rsm} \partial_z  + \frac{1}{ \sqrt{2} }\Mo^{\Rsm\Lsm}\partial_{\Po^\Rsm} \partial_z - \frac{\Delta_\beta}{18}   \partial_{\Po^\Rsm}^2    \partial^2_z\Bigr)\Bigr) |\Vb\rangle   = 0\,,
\eeq
while the equation \rf{04072018-man02-20} leads to the following decoupled equations for the  holomorphic and anti-holomorphic ket-vectors $|V\rangle = V|0\rangle$, $|\Vb\rangle = \Vb|0\rangle$:
%:
%
\beq
\label{07072018-man02-18} && \hspace{-1.2cm} \Bigl(-  \Po^\Lsm \No_\beta  - \!\Mo^{\Rsm\Lsm} \Po^\Lsm + \!\frac{\sqrt{2} \beta}{3} \MM^{\Rsm\Lsm}\partial_z + \!\frac{\sqrt{2} \Delta_\beta}{36} N_{\Po^\Lsm} \partial_z +  \!\frac{\sqrt{2}\Delta_\beta}{36} \Mbf^{\Rsm\Lsm} \partial_z
+ \!\frac{\betach}{54} \partial_{\Po^\Lsm}  \partial_z^2\Bigr)|V\rangle  = 0 \,,\qquad
\\
\label{07072018-man02-19} && \hspace{-1.2cm} \Bigl(- \Po^\Rsm \No_\beta  + \! \Mo^{\Rsm\Lsm} \Po^\Rsm  + \!\frac{\sqrt{2} \beta}{3} \MM^{\Rsm\Lsm}\partial_z  - \!\frac{\sqrt{2} \Delta_\beta}{36} N_{\Po^\Rsm} \partial_z +  \!\frac{\sqrt{2} \Delta_\beta}{36}  \Mbf^{\Rsm\Lsm} \partial_z
+ \!\frac{\betach}{54}  \partial_{\Po^\Rsm} \partial_z^2\Bigr)|\Vb\rangle  = 0 \,,\qquad
\eeq
where $\Delta_\beta$, $\beta$, $\betach$, $\No_\beta$ are given in \rf{16072018-man02-11}-\rf{16072018-man02-15} and we use the notation
\beq
\label{07072018-man02-08} && \hspace{-0.8cm} \Mbf^{\Rsm\Lsm} =  \sum_{a=1,2,3} M_a^{\Rsm\Lsm}\,, \qquad  \Mo^{\Rsm\Lsm} =  \frac{1}{3} \sum_{a=1,2,3} \betach_a M_a^{\Rsm\Lsm}\,,
\qquad  \MM^{\Rsm\Lsm} =  \sum_{a=1,2,3} \frac{1}{\beta_a} M_a^{\Rsm\Lsm}\,,\qquad
\\
\label{07072018-man02-07} && \hspace{-0.8cm} \Jbf^{\Rsm\Lsm} =  \Po^\Rsm \partial_{\Po^\Rsm} - \Po^\Lsm \partial_{\Po^\Lsm} +  \Mbf^{\Rsm\Lsm}\,,
\hspace{1cm} M_a^{\Rsm\Lsm} = \alpha_a^\Rsm \alphab_a^\Lsm - \alpha_a^\Lsm \alphab_a^\Rsm\,.
\eeq

\noindent {\bf Step 8 }. Our aim at this step is to find solution to equations for vertices $V$, $\Vb$ given in \rf{14072018-man02-02},\rf{14072018-man02-03} and \rf{07072018-man02-05}-\rf{07072018-man02-19}. The equations for the vertices $V$ and $\Vb$ are similar. Therefore, to avoid the repetitions, we focus on the study of the holomorphic vertex $V$.
To this end we introduce a new vertex $|V_0\rangle$ which is related to the vertex $|V\rangle$ by the following invertible transformation:
\beq
\label{07072018-man02-10} && |V\rangle = U |V_0\rangle\,,
\\
\label{07072018-man02-11}  && |V_0\rangle = V_0|0\rangle\,, \qquad V_0 = V_0(\Po^\Lsm,\beta_a,z,\alpha_a)\,,
\eeq
where the operator $U$ is defined by the relations
\beq
\label{07072018-man02-14} && U = \overleftarrow{T} \exp\bigl( \int_0^1 d\tau\, u_\tau\bigr)\,,
\\
\label{07072018-man02-15} && u_t = - \frac{1}{N_z + 2}\Bigl( \frac{1}{\sqrt{2}} \No_\beta \partial_{\Po^\Lsm} \partial_z  - \frac{1}{ \sqrt{2} }\Mo^{\Rsm\Lsm}\partial_{\Po^\Lsm} \partial_z + \frac{t\Delta_\beta}{18}   \partial_{\Po^\Lsm}^2    \partial^2_z \Bigr)\,,
\eeq
while $\Mo^{\Rsm\Lsm}$, $\No_\beta$, $\Delta_\beta$ are defined in \rf{16072018-man02-08},\rf{16072018-man02-11},\rf{16072018-man02-15}.
Remarkable feature of the transformation \rf{07072018-man02-10}  is that, in terms of the ket-vector $|V_0\rangle$, equation \rf{07072018-man02-05} takes the following simple form:
\be \label{07072018-man02-16}
\Jbf^{\Rsm\Lsm} |V_0\rangle = 0\,.
\ee
Equation \rf{07072018-man02-16} coincides with the one for cubic vertex of massless fields in flat space. Using the expression for the operator $\Jbf^{\Rsm\Lsm}$ \rf{07072018-man02-07}, we see that solution to equation \rf{07072018-man02-16} is given by
\be \label{07072018-man02-17}
V_0(\Po^\Lsm,\beta_a,z,\alpha_a) = (\Po^\Lsm)^{\Mbf^{\Rsm\Lsm}} V_0'\,, \qquad V_0'=V_0'(\beta_a,z,\alpha_a)\,.
\ee
Thus, the dependence of the vertex $V_0$ on the momentum $\Po^\Lsm$ is completely fixed.

Using solution in \rf{07072018-man02-17}, we now analyse equations \rf{14072018-man02-02}
and \rf{07072018-man02-18}. To this end we note that plugging \rf{07072018-man02-17} into \rf{07072018-man02-18}, we find that equation \rf{07072018-man02-18} leads to the following simple equation for the vertex $|V_0'\rangle = V_0'|0\rangle$ \rf{07072018-man02-17}:
\be \label{07072018-man02-20}
\bigl(\No_\beta  + \Mo^{\Rsm\Lsm}  \bigr) |V_0'\rangle  = 0\,,
\ee
while, plugging  \rf{07072018-man02-17} into \rf{14072018-man02-02}, we obtain the equations
\beq
\label{07072018-man02-21} && \bigl( \Mbf^{\Rsm\Lsm} + \sum_{a=1,2,3} \beta_a\partial_{\beta_a} \bigr) |V_0'\rangle = 0\,,
\\
\label{07072018-man02-22} && \bigl( N_z  - \Mbf^{\Rsm\Lsm}  + 1 \bigr) |V_0'\rangle = 0 \,.
\eeq
Note that equations \rf{07072018-man02-20},\rf{07072018-man02-21} coincide with the ones for cubic vertex of massless fields in flat space. Introducing a new vertex $V_0''$ by the relation
\be \label{07072018-man02-23}
V_0' =  \frac{ z^{\Mbf^{\Rsm\Lsm}-1} }{ \beta_1^{M_1^{\Rsm\Lsm}}\beta_2^{M_2^{\Rsm\Lsm}}\beta_3^{M_3^{\Rsm\Lsm}} } V_0''\,,\qquad V_0''=V_0''(\beta_a,z,\alpha_a)\,,
\ee
we find that, in terms of the new vertex $V_0''$, equations \rf{07072018-man02-20}-\rf{07072018-man02-22} take the form
\beq
\label{07072018-man02-24} && \No_\beta V_0''  = 0\,,
\\
\label{07072018-man02-25} && \sum_{a=1,2,3} \beta_a\partial_{\beta_a} V_0'' = 0\,,
\\
\label{07072018-man02-26} &&  N_z V_0'' = 0 \,.
\eeq
From equations \rf{07072018-man02-24},\rf{07072018-man02-25}, we learn that the vertex $V_0''$ is independent of the momenta $\beta_1$,$\beta_2$,$\beta_3$, while equation \rf{07072018-man02-26} tells us that the vertex $V_0''$ is independent of the coordinate $z$. In other words, equations \rf{07072018-man02-24}-\rf{07072018-man02-26} imply that vertex $V_0''$ \rf{07072018-man02-23} depends only on the oscillators,
\be \label{07072018-man02-27}
V_0'' = C(\alpha_a)\,, \qquad \alpha_a = \alpha_a^\Rsm,\alpha_a^\Lsm\,,
\ee
where, in \rf{07072018-man02-27}, we recall that the argument $\alpha_a$ stands for the oscillators $\alpha_a^\Rsm$, $\alpha_a^\Lsm$.

Thus, we exhaust all equations we imposed on the cubic vertex. We find that the general solution for the cubic vertex is governed by the vertex $C(\alpha_a)$ which depends only on the oscillators. Collecting relations \rf{07072018-man02-17}, \rf{07072018-man02-23}, \rf{07072018-man02-27}, we see that general solution for the vertex $V_0$ entering the cubic interaction vertex in \rf{07072018-man02-03}, \rf{07072018-man02-10} takes the following form:
\be \label{07072018-man02-27-a1}
V_0 =   \frac{ (z\Po^\Lsm)_{\vphantom{5pt}}^{\Mbf^{\Rsm\Lsm}} }{z\, \beta_1^{ M_1^{\Rsm\Lsm} } \beta_2^{M_2^{\Rsm\Lsm} } \beta_3^{ M_3^{\Rsm\Lsm} } } C(\alpha_a)\,.
\ee
Note that the full expression for the holomorphic vertex $V$ entering cubic vertex $p_\smp3^-$ \rf{07072018-man02-03} is obtained by using the relations given in \rf{07072018-man02-10}-\rf{07072018-man02-15}.

The procedure above described can be used for the derivation of the explicit representation for the anti-holomorphic vertex $\Vb$. In next Section, we summarize our results for both the holomorphic and anti-holomorphic vertices.

%%%%%%%%%%%%%%%%%%%%%%%%%%%%%%%%%%%%%%%%%%%%%%%%%%%%%%%%%%%%%%%%%%%%%%%%%%%%%%%%%%%%%%%%%%%
%%%%%%%%%%%%%%%%%%%%%%%%%%%%%%%%%%%%%%%%%%%%%%%%%%%%%%%%%%%%%%%%%%%%%%%%%%%%%%%%%%%%%%%%%%%
\newsection{ \large Cubic interaction vertex for massless AdS fields }\label{sec-06}
%%%%%%%%%%%%%%%%%%%%%%%%%%%%%%%%%%%%%%%%%%%%%%%%%%%%%%%%%%%%%%%%%%%%%%%%%%%%%%%%%%%%%%%%%%
%%%%%%%%%%%%%%%%%%%%%%%%%%%%%%%%%%%%%%%%%%%%%%%%%%%%%%%%%%%%%%%%%%%%%%%%%%%%%%%%%%%%%%%%%%

We now summarize our result for cubic interaction vertex we obtained in the previous sections. For the reader's convenience, we present the representation for the cubic vertex in terms of generating functions as well as the representation for the cubic vertex in terms of the component fields.

\noindent {\bf Generating form of cubic vertex}. Let us refer to cubic vertex that describes interaction of spin-$s_1$, spin-$s_2$, and spin-$s_3$ massless fields as $s_1$-$s_2$-$s_3$ cubic vertex. We note then that solution for cubic interaction vertex we found is given by
\be  \label{09072018-man02-01}
p_\smp3^- = \frac{C_\scalarrm^{000}}{z} + V + \Vb\,,
\ee
where the $C_\scalarrm^{000}$ term describes the 0-0-0 cubic vertex, while the holomorphic vertex $V$ and anti-holomorphic vertex $\Vb$ describe the $s_1$-$s_2$-$s_3$ cubic vertices when $s_1+s_2+s_3>0$, $s_1\geq 0$, $s_2\geq 0$, $s_3 \geq 0$. Explicit expressions for the holomorphic and anti-holomorphic vertices $V$, $\Vb$ are given by
\beq
\label{09072018-man02-02} && V =   U V_0\,, \qquad \Vb =   \Ub \Vb_0\,,
\\
\label{09072018-man02-03} && V_0 =   \frac{ (z\Po^\Lsm)_{\vphantom{5pt}}^{\Mbf^{\Rsm\Lsm}} }{z\, \beta_1^{ M_1^{\Rsm\Lsm} } \beta_2^{M_2^{\Rsm\Lsm} } \beta_3^{ M_3^{\Rsm\Lsm} } } C\,,
\\
\label{09072018-man02-04} && \Vb_0 =   \frac{ (z\Po^\Rsm)_{\vphantom{5pt}}^{-\Mbf^{\Rsm\Lsm}} }{z\, \beta_1^{- M_1^{\Rsm\Lsm} } \beta_2^{ -M_2^{\Rsm\Lsm} } \beta_3^{ -M_3^{\Rsm\Lsm} } } \Cb\,,
\\
\label{09072018-man02-05} && C  = \sum_{\lambda_1,\lambda_2,\lambda_3=-\infty }^\infty   C^{\lambda_1\lambda_2\lambda_3}
\frac{ \alpha_{\Hsm,1}^{\lambda_1} \alpha_{\Hsm,2}^{\lambda_2} \alpha_{\Hsm,3}^{\lambda_3}}{\sqrt{|\lambda_1|!|\lambda_2|!|\lambda_3|!}}\,,
\\
\label{09072018-man02-06} && \Cb = \sum_{ \lambda_1,\lambda_2,\lambda_3 = -\infty }^\infty   \Cb^{\lambda_1\lambda_2\lambda_3}
\frac{ \alpha_{\Hsm,1}^{\lambda_1} \alpha_{\Hsm,2}^{\lambda_2} \alpha_{\Hsm,3}^{\lambda_3}}{\sqrt{|\lambda_1|!|\lambda_2|!|\lambda_3|!}}\,,
\eeq
where the symbol $\alpha_\Hsm^\lambda$ is defined in \rf{27062018-man02-14}, while the spin operators $\Mbf^{\Rsm\Lsm}$, $M_a^{\Rsm\Lsm}$ are defined below in \rf{09072018-man02-14},\rf{09072018-man02-15}.
Definition of the momenta $\Po^\Lsm$, $\Po^\Rsm$ is given in \rf{30062018-man02-36}, \rf{03072018-man02-16}, \rf{07072018-man02-01}. In \rf{09072018-man02-05}, \rf{09072018-man02-06} and below, the quantities $C^{\lambda_1\lambda_2\lambda_3}$ and $\Cb^{\lambda_1\lambda_2\lambda_3}$ are constant parameters. These constant parameters are freedom of our solution. As we shall see below, the $C^{\lambda_1\lambda_2\lambda_3}$ is a coupling constant describing interaction of three massless fields having helicities $-\lambda_1$, $-\lambda_2$, $-\lambda_3$ that satisfy the restriction $\lambda_1+\lambda_2+\lambda_3\geq 1$, while  $\Cb^{\lambda_1\lambda_2\lambda_3}$ is a coupling constant describing interaction of three massless fields having helicities $-\lambda_1$, $-\lambda_2$, $-\lambda_3$ that satisfy the restriction $-\lambda_1-\lambda_2-\lambda_3\geq 1$. In other words, the coupling constants $C^{\lambda_1\lambda_2\lambda_3}$ and $\Cb^{\lambda_1\lambda_2\lambda_3}$ satisfy the restrictions
\beq
\label{09072018-man02-07} && C^{\lambda_1\lambda_2\lambda_3} \ne 0\,, \hspace{1cm} \hbox{for } \quad \lambda_1+\lambda_2+\lambda_3 \geq 1 \,,
\\
\label{09072018-man02-08} && \Cb^{\lambda_1\lambda_2\lambda_3} \ne 0\,, \hspace{1cm} \hbox{for } -\lambda_1-\lambda_2-\lambda_3 \geq  1\,.
\eeq
Operators $U$ and $\Ub$ appearing in \rf{09072018-man02-02} are given by
\be \label{09072018-man02-09}
U  =  \overleftarrow{T}  \exp\Bigl(\int_0^1 d\tau u_\tau \Bigr)\,, \qquad \Ub = \overleftarrow{T}  \exp\Bigl(\int_0^1 d\tau \ub_\tau\Bigr)\,,
\ee
where operators $u_t$, $\ub_t$ are defined by the relations
\beq
\label{09072018-man02-10} && u_t  =   \frac{1}{\sqrt{2}} \bigl(-\No_\beta  + \Mo^{\Rsm\Lsm}\bigr) Y - \frac{t\Delta_\beta}{18}    Y^2 N_{\Po^\Lsm}\,,
\\
\label{09072018-man02-11} && \ub_t =   \frac{1}{\sqrt{2}} \bigl(\No_\beta  + \Mo^{\Rsm\Lsm}\bigr) \Yb - \frac{t\Delta_\beta}{18}    \Yb^2 N_{\Po^\Rsm}\,,
\\
\label{09072018-man02-12} && Y = \frac{1}{N_z+2}\partial_z \partial_{\Po^\Lsm}\,, \qquad \Yb = \frac{1}{N_z+2}\partial_z \partial_{\Po^\Rsm} \,,
\\
\label{09072018-man02-12-a1} && N_{\Po^\Lsm} = \Po^\Lsm \partial_{\Po^\Lsm}\,, \hspace{1.7cm} N_{\Po^\Rsm} = \Po^\Rsm \partial_{\Po^\Rsm}\,, \qquad N_z= z\partial_z\,,
\eeq
while the quantities $\Mo^{\Rsm\Lsm}$, $\No_\beta$, $\Delta_\beta$ are defined in \rf{16072018-man02-08},\rf{16072018-man02-11},\rf{16072018-man02-15} in Appendix A.
We recall also that the spin  operators $M_a^{\Rsm\Lsm}$ and various quantities constructed out of $M_a^{\Rsm\Lsm}$ and the momenta $\beta_1,\beta_2,\beta_3$ are defined by the relations
\beq
\label{09072018-man02-14}  && \hspace{-1cm} M_a^{\Rsm\Lsm} = \alpha_a^\Rsm \alphab_a^\Lsm - \alpha_a^\Lsm \alphab_a^\Rsm\,,
\\
\label{09072018-man02-15} && \hspace{-1cm} \Mbf^{\Rsm\Lsm} =  \sum_{a=1,2,3} M_a^{\Rsm\Lsm}\,, \qquad \Mo^{\Rsm\Lsm} =  \frac{1}{3} \sum_{a=1,2,3} \betach_a M_a^{\Rsm\Lsm}\,, \qquad \MM^{\Rsm\Lsm} =  \sum_{a=1,2,3} \frac{1}{\beta_a} M_a^{\Rsm\Lsm}\,.\qquad
\eeq
Acting with the operator $\No_\beta$ on the quantities given in \rf{09072018-man02-15}, we get relations which can be helpful for the computation of the $\tau$-ordered exponentials in \rf{09072018-man02-09},
\beq
\label{09072018-man02-16}  && \No_\beta \Mbf^{\Rsm\Lsm}  =  0\,,
\\
\label{09072018-man02-17} && \No_\beta \Mo^{\Rsm\Lsm} = \frac{\Delta_\beta}{18} \Mbf^{\Rsm\Lsm} +  \frac{\beta}{3} \MM^{\Rsm\Lsm}\,,
\\
\label{09072018-man02-18} && \No_\beta \MM^{\Rsm\Lsm} = \frac{\betach}{9\beta} \Mbf^{\Rsm\Lsm} + \frac{\Delta_\beta}{3\beta} \Mo^{\Rsm\Lsm}\,.
\eeq

Note that, when writing the solution for $C$ and $\Cb$ in \rf{09072018-man02-05},\rf{09072018-man02-06}, we use constraint \rf{27062018-man02-10}. This constraint implies that the cubic vertex does not depend on the $(\alpha_a^\Rsm\alpha_a^\Lsm)^{n_a}$-terms for $n_a>0$.

Our cubic vertices for massless fields in $AdS_4$ are simply related to cubic vertices for massless field in flat space. Namely, if we multiply the holomorphic vertex $V_0$ \rf{09072018-man02-03} by the factor $z^{1- \Mbf^{\Rsm\Lsm}}$, then we get holomorphic vertex for massless fields in flat space. Also we note that, if we multiply the anti-holomorphic vertex $\Vb_0$ \rf{09072018-man02-03} by the factor $z^{1+ \Mbf^{\Rsm\Lsm}}$, then we get anti-holomorphic vertex for massless fields in flat space. This is to say that, by module of the overall factor $Uz^{\Mbf^{\Rsm\Lsm}-1}$, the holomorphic cubic vertex $V$ \rf{09072018-man02-02} for massless fields in $AdS_4$ coincides with the holomorphic cubic vertex for massless fields in flat space, while, by module of the overall factor $\Ub z^{-\Mbf^{\Rsm\Lsm}-1}$, the anti-holomorphic cubic vertex $\Vb$ \rf{09072018-man02-02} for massless fields in $AdS_4$ coincides with the anti-holomorphic cubic vertex for massless fields in flat space.
For massless fields in the four-dimensional flat space, the complete list of cubic vertices was obtained in Ref.\cite{Bengtsson:1986kh}. Thus we see that all cubic vertices for massless fields in flat space obtained in Ref.\cite{Bengtsson:1986kh} have their counterparts in AdS space.

\noindent {\bf Alternative representation for operators $U$, $\Ub$}. Operator $U$  \rf{09072018-man02-09} is realized as differential operators with respect to the momenta $\Po^\Lsm$, $\beta_1$, $\beta_2$, $\beta_3$, and the coordinate $z$, while, operator $\Ub$  \rf{09072018-man02-09} is realized as differential operators with respect to the momenta $\Po^\Rsm$, $\beta_1,\beta_2,\beta_3$, and the coordinates $z$.
Note that the derivatives of the momenta $\beta_1$, $\beta_2$, $\beta_3$ enter the operators $U$, $\Ub$ through the operator $\No_\beta$. Using equation \rf{07072018-man02-18}, we see that on space of the vertex $V$, the operator $\No_\beta$ can be replaced by differential operator with respect to momentum $\Po^\Lsm$ and the coordinate $z$, while, using \rf{07072018-man02-19}, we see that, on space of the vertex $\Vb$, the operator $\No_\beta$ can be replaced by differential operator with respect to the momentum $\Po^\Rsm$, and the coordinate $z$.
Doing so, we get the following alternative representation for the holomorphic and anti-holomorphic vertices $V$, $\Vb$,
\be \label{09072018-man02-19}
V  =   U_\Po^\vph  V_0\,, \qquad \Vb =   \Ub_\Po^\vph \Vb_0\,,
\ee
where $V_0$, $\Vb_0$ are given in \rf{09072018-man02-03},\rf{09072018-man02-04}, while the operators $U_\Po$ and $\Ub_\Po$ are defined by the relations
\beq \label{09072018-man02-20}
&& \hspace{-1cm} U_\Po  = \overleftarrow{T} \exp\Bigl(\int_0^1 d\tau u_{\Po,\tau}^\vph  \Bigr)\,, \qquad \Ub_\Po =  \overleftarrow{T}  \exp\Bigl(\int_0^1 d\tau \ub_{\Po,\tau}^\vph \Bigr),
\\
\label{09072018-man02-21}  && u_{\Po,t}^\vph =  \sqrt{2}\Mo^{\Rsm\Lsm} Y -  \frac{t\beta}{3} \MM^{\Rsm\Lsm} Y^2  - \frac{t\Delta_\beta}{36} \Mbf^{\Rsm\Lsm} Y^2 -  \frac{t \Delta_\beta}{12}   Y^2 N_{\Po^\Lsm} - \frac{t^2\sqrt{2}\betach}{108} Y^3 N_{\Po^\Lsm}\,,\qquad
\\
\label{09072018-man02-22} && \ub_{\Po,t}^\vph =  \sqrt{2}\Mo^{\Rsm\Lsm} \Yb +  \frac{t\beta}{3} \MM^{\Rsm\Lsm} \Yb^2  + \frac{t\Delta_\beta}{36} \Mbf^{\Rsm\Lsm} \Yb^2 -  \frac{t \Delta_\beta}{12}   \Yb^2 N_{\Po^\Rsm} + \frac{t^2\sqrt{2}\betach}{108} \Yb^3 N_{\Po^\Rsm}\,,\qquad
\eeq
and $Y$, $\Yb$ are given in \rf{09072018-man02-12},\rf{09072018-man02-14}. For the definition of various quantities appearing in \rf{09072018-man02-21},\rf{09072018-man02-22}, see relations \rf{16072018-man02-07}-\rf{16072018-man02-14} in Appendix A.

Expressions \rf{09072018-man02-20},\rf{09072018-man02-21} provide the realization of the operator $U$ in terms of differential operators with respect to the momentum $\Po^\Lsm$ and the coordinate $z$, while expressions \rf{09072018-man02-20},\rf{09072018-man02-22} provide the realization of the operator $\Ub$ in terms of differential operators with respect to the momentum $\Po^\Rsm$ and the coordinate $z$.

\noindent {\bf Component form of cubic vertex}. Massless spin-$s$, $s>0$, field in $AdS_4$ is described by two complex-valued fields $\phi_\lambda$, $\lambda=\pm s$ \rf{27062018-man02-12}. Plugging vertex \rf{09072018-man02-01} into \rf{30062018-man02-02} and using the representation of the bra-vector $\langle\phi|$ \rf{27062018-man02-14-a1} and relations \rf{30062018-man02-05}, \rf{30062018-man02-06}, we can work out an explicit representation for the cubic Hamiltonian \rf{30062018-man02-02} in terms of the component complex-valued fields $\phi_\lambda^\dagger$. Computation of the component form of the cubic Hamiltonian is simplified by noticing the following relations for the spin operators $M_a^{\Rsm\Lsm}$:
\be \label{09072018-man02-23}
M_a^{\Rsm\Lsm} \alpha_{\Hsm,a}^{\lambda_a} |0\rangle = \lambda_a \alpha_{\Hsm,a}^{\lambda_a} |0\rangle\,, \qquad a=1,2,3\,.
\ee
Using relations \rf{09072018-man02-23}, we get the following component form for the cubic vertex:
\be \label{09072018-man02-23-a1}
\langle \phi_1|\langle \phi_2|\langle \phi_3| p_\smp3^-\rangle  =  \Phi_{000}^\dagger V_\scalarrm^{000} + \sum_{\lambda_1,\lambda_2,\lambda_3=-\infty}^\infty \Phi_{\lambda_1,\lambda_2,\lambda_3}^\dagger \bigl( V^{\lambda_1\lambda_2\lambda_3}  +  \Vb^{\lambda_1\lambda_2\lambda_3} \bigr)\,,
\ee
where we use the notation
\beq
\label{09072018-man02-24} && V_\scalarrm^{000} \equiv  \frac{C_\scalarrm^{000}}{z}\,,
\\
\label{09072018-man02-25} && V^{\lambda_1\lambda_2\lambda_3} \equiv  C^{\lambda_1\lambda_2\lambda_3} U^{\lambda_1\lambda_2\lambda_3}  \frac{(z\Po^\Lsm)_{\vphantom{5pt}}^{\lambda_1+\lambda_2+\lambda_3}}{z\, \beta_1^{\lambda_1} \beta_2^{\lambda_2} \beta_3^{\lambda_3} }\,,
\\
\label{09072018-man02-26} && \Vb^{\lambda_1\lambda_2\lambda_3} \equiv \Cb^{\lambda_1\lambda_2\lambda_3} \Ub^{\lambda_1\lambda_2\lambda_3} \frac{(z\Po^\Rsm)_{\vphantom{5pt}}^{-\lambda_1-\lambda_2-\lambda_3}}{z\, \beta_1^{-\lambda_1} \beta_2^{-\lambda_2} \beta_3^{-\lambda_3} }\,,
\\
\label{09072018-man02-27} && \Phi_{\lambda_1,\lambda_2,\lambda_3}^\dagger \equiv \phi_{\lambda_1}^\dagger(p_1,z_1)\phi_{\lambda_2}^\dagger(p_2,z_2)\phi_{\lambda_3}^\dagger(p_3,z_3)\,,
\eeq
and the coupling constants $C^{\lambda_1\lambda_2\lambda_3}$ and $\Cb^{\lambda_1\lambda_2\lambda_3}$ should satisfy the restrictions in \rf{09072018-man02-07},\rf{09072018-man02-08}.

Operators $U^{\lambda_1\lambda_2\lambda_3}$ and $\Ub^{\lambda_1\lambda_2\lambda_3}$ appearing in \rf{09072018-man02-25}, \rf{09072018-man02-26} are obtained from the respective operators $U$ and $\Ub$ by using the replacement
implied by the relations \rf{09072018-man02-23}
\beq
\label{09072018-man02-28} && U^{\lambda_1\lambda_2\lambda_3} = U|_{\Mbf^{\Rsm\Lsm}\rightarrow \Mbf_\lambda^{\Rsm\Lsm},\Mo^{\Rsm\Lsm}\rightarrow \Mo_\lambda^{\Rsm\Lsm},\MM^{\Rsm\Lsm}\rightarrow \MM_\lambda^{\Rsm\Lsm}}\,,
\\
\label{09072018-man02-29} && \Ub^{\lambda_1\lambda_2\lambda_3} = \Ub|_{\Mbf^{\Rsm\Lsm}\rightarrow \Mbf_\lambda^{\Rsm\Lsm},\Mo^{\Rsm\Lsm}\rightarrow \Mo_\lambda^{\Rsm\Lsm},\MM^{\Rsm\Lsm}\rightarrow \MM_\lambda^{\Rsm\Lsm}}\,,
\eeq
where we use the notation
\be \label{09072018-man02-30}
\Mbf_\lambda^{\Rsm\Lsm} = \sum_{a=1,2,3} \lambda_a\,,
\qquad  \Mo_\lambda^{\Rsm\Lsm} = \frac{1}{3}\sum_{a=1,2,3} \betach_a \lambda_a\,,
\qquad \MM_\lambda^{\Rsm\Lsm} =  \sum_{a=1,2,3} \frac{\lambda_a}{\beta_a}\,.
\ee

Alternative representation for vertices $V^{\lambda_1\lambda_2\lambda_3}$, $\Vb^{\lambda_1\lambda_2\lambda_3}$ \rf{09072018-man02-25}, \rf{09072018-man02-26} associated with the ones in \rf{09072018-man02-19} can be obtained by making on r.h.s. in \rf{09072018-man02-25}, \rf{09072018-man02-26} the following replacements:
\be  \label{09072018-man02-31}
U^{\lambda_1\lambda_2\lambda_3}\rightarrow U_\Po^{\lambda_1\lambda_2\lambda_3}\,, \qquad \Ub^{\lambda_1\lambda_2\lambda_3}\rightarrow \Ub_\Po^{\lambda_1\lambda_2\lambda_3}\,,
\ee
where operators $U_\Po^{\lambda_1\lambda_2\lambda_3}$ and $\Ub_\Po^{\lambda_1\lambda_2\lambda_3}$  are obtained from the respective operators $U_\Po$ and $\Ub_\Po$ by using the replacement
implied by the relations \rf{09072018-man02-23},
\beq
\label{09072018-man02-32} && U_\Po^{\lambda_1\lambda_2\lambda_3} = U_\Po|_{\Mbf^{\Rsm\Lsm}\rightarrow \Mbf_\lambda^{\Rsm\Lsm},\Mo^{\Rsm\Lsm}\rightarrow \Mo_\lambda^{\Rsm\Lsm},\MM^{\Rsm\Lsm}\rightarrow \MM_\lambda^{\Rsm\Lsm}}\,,
\\
\label{09072018-man02-33} && \Ub_\Po^{\lambda_1\lambda_2\lambda_3} = \Ub_\Po|_{\Mbf^{\Rsm\Lsm}\rightarrow \Mbf_\lambda^{\Rsm\Lsm},\Mo^{\Rsm\Lsm}\rightarrow \Mo_\lambda^{\Rsm\Lsm},\MM^{\Rsm\Lsm}\rightarrow \MM_\lambda^{\Rsm\Lsm}}\,.
\eeq

The following remarks are in order.

\noindent \ibf) Field $\phi_\lambda$ \rf{27062018-man02-12} describes a massless field having the helicity equal to $\lambda$, while the hermitian-conjugated field $\phi_\lambda^\dagger$  \rf{27062018-man02-14-a1} describes a massless field having the opposite helicity equal to $-\lambda$. Note that it is the fields $\phi_{\lambda_1}^\dagger, \phi_{\lambda_2}^\dagger, \phi_{\lambda_3}^\dagger$ that enter the cubic vertex \rf{09072018-man02-23-a1}, \rf{09072018-man02-27}. Therefore the cubic vertex \rf{09072018-man02-23-a1} describes interaction of three fields having the helicities $-\lambda_1$, $-\lambda_2$, $-\lambda_3$.

\noindent \iibf) By definition, vertex $V^{\lambda_1\lambda_2\lambda_3}$ \rf{09072018-man02-25} is polynomial in  $\Po^\Lsm$, while vertex $\Vb^{\lambda_1\lambda_2\lambda_3}$ \rf{09072018-man02-26} is polynomial in  $\Po^\Rsm$. Below we demonstrate that the vertex $V^{\lambda_1\lambda_2\lambda_3}$ \rf{09072018-man02-25} does not involve terms of zero-order in $\Po^\Lsm$, while vertex $\Vb^{\lambda_1\lambda_2\lambda_3}$ \rf{09072018-man02-26}
does not involve terms of zero-order in $\Po^\Rsm$. Taking this into account, we get the restriction $\lambda_1 + \lambda_2 + \lambda_3 \geq 1$ for the vertex $V^{\lambda_1\lambda_2\lambda_3}$  and the restriction $-\lambda_1 - \lambda_2 - \lambda_3 \geq 1$ for the vertex $\Vb^{\lambda_1\lambda_2\lambda_3}$.
Thus, as we stated earlier, the $C^{\lambda_1\lambda_2\lambda_3}$ is a coupling constant describing interaction of three massless fields having helicities $-\lambda_1$, $-\lambda_2$, $-\lambda_3$ that satisfy the restriction $\lambda_1+\lambda_2+\lambda_3\geq 1$, while  $\Cb^{\lambda_1\lambda_2\lambda_3}$ is a coupling constant describing interaction of three massless fields having helicities $-\lambda_1$, $-\lambda_2$, $-\lambda_3$ that satisfy the restriction $-\lambda_1-\lambda_2-\lambda_3\geq 1$.

\noindent \iiibf) Using expressions for the operators $u_t$, $\ub_t$ given in \rf{09072018-man02-10}-\rf{09072018-man02-12}, we see that powers of the momenta $\Po^\Lsm$, $\Po^\Rsm$ decrease upon acting on the vertices $V_0$, $\Vb_0$ with the operators $u_t$, $\ub_t$. This implies that maximal number of powers of momentum $\Po^\Lsm$ appearing in the vertex $V^{\lambda_1\lambda_2\lambda_3}$ \rf{09072018-man02-25} is equal to $\lambda_1 + \lambda_2 + \lambda_3$, while a maximal number of powers of the momentum $\Po^\Rsm$ appearing in the vertex $\Vb^{\lambda_1\lambda_2\lambda_3}$ \rf{09072018-man02-26} is equal to $-\lambda_1 - \lambda_2 - \lambda_3$. Taking this into account we now note that vertices \rf{09072018-man02-25}, \rf{09072018-man02-26} have the following expansions in the momenta $\Po^\Lsm$, $\Po^\Rsm$, and the coordinate $z$
\beq
\label{09072018-man02-34} && V^{\lambda_1\lambda_2\lambda_3} = \sum_{n=1}^{\lambda_1+\lambda_2+\lambda_3} (\Po^\Lsm)^n z^{n-1} V_n^{\lambda_1\lambda_2\lambda_3}\,, \hspace{1.2cm} \hbox{ for }\  \lambda_1+\lambda_2+\lambda_3\geq 1\,,\qquad
\\
\label{09072018-man02-35} && \Vb^{\lambda_1\lambda_2\lambda_3} = \sum_{n=1}^{-\lambda_1-\lambda_2-\lambda_3} (\Po^\Rsm)^n z^{n-1} \Vb_n^{\lambda_1\lambda_2\lambda_3}\,, \hspace{1cm} \hbox{ for }\  -\lambda_1-\lambda_2-\lambda_3\geq 1\,,\qquad
\eeq
where the expansion coefficients $V_n^{\lambda_1\lambda_2\lambda_3}$ and $\Vb_n^{\lambda_1\lambda_2\lambda_3}$ depend only on the momenta $\beta_1,\beta_2,\beta_3$ and the helicities $\lambda_1,\lambda_2,\lambda_3$.
Note that in the expansions \rf{09072018-man02-34} and \rf{09072018-man02-35} there are no terms of zero-order in  $\Po^\Lsm$ and $\Po^\Rsm$ respectively.
Absence of the terms of zero-order in $\Po^\Lsm$ and $\Po^\Rsm$ in the respective expansions \rf{09072018-man02-34} and \rf{09072018-man02-35} can be verified by using \rf{09072018-man02-25},  \rf{09072018-man02-26} and the following respective relations for the operators $u_t$, $\ub_t$:
\beq
\label{09072018-man02-36} u_{\tau_1} \ldots u_{\tau_n} (\Po^\Lsm)^m z^{m-1} \Bigr|_{\Po^\Lsm=0}=0\,, \hspace{1cm} \hbox{ for } n\geq 0\,, \quad m\geq 1\,,
\\
\label{09072018-man02-37} \ub_{\tau_1} \ldots \ub_{\tau_n} (\Po^\Rsm)^m z^{m-1} \Bigr|_{\Po^\Rsm=0}=0\,, \hspace{1cm} \hbox{ for } n\geq 0\,, \quad m\geq 1\,,
\eeq
where operators $u_t$, $\ub_t$ are defined in \rf{09072018-man02-10}-\rf{09072018-man02-12}.

\noindent \ivbf) The coupling constants $C_\scalarrm^{000}$, $C^{\lambda_1\lambda_2\lambda_3}$, $\Cb^{\lambda_1\lambda_2\lambda_3}$ appearing in \rf{09072018-man02-24}-\rf{09072018-man02-26} are dimensionless. Obviously, these coupling constants turn out to be dimensionless in view of the particular powers of the radial coordinate $z$ appearing in vertices \rf{09072018-man02-24}-\rf{09072018-man02-26} .

\noindent \vbf) For the case of internal algebra $o(\Nsf)$, incorporation of a internal symmetry into the theory of massless AdS fields can be realized in the same way as for massless fields in flat space. Namely, first, in place of fields $\phi_\lambda$, we introduce fields $\phi_\lambda^{\asf\bsf}$, where the indices $\asf,\bsf$ are the matrix indices of the $o(\Nsf)$ algebra, $\asf,\bsf=1,\ldots,\Nsf$. By definition, the fields satisfy the relation $\phi_\lambda^{\asf\bsf} = (-)^\lambda\phi_\lambda^{\bsf\asf}$. Hermicity property of fields are as follows $(\phi_\lambda^{\asf\bsf}(p,z))^\dagger = \phi_{-\lambda}^{\asf\bsf}(-p,z)$. Second, in scalar products, the expressions $\phi_\lambda^\dagger \phi_\lambda$ are replaced by  $\phi_\lambda^{\asf\bsf\dagger} \phi_\lambda^{\asf\bsf}$,
while, in cubic vertices, the expressions $\phi_{\lambda_1}^\dagger
\phi_{\lambda_2}^\dagger \phi_{\lambda_3}^\dagger$ are replaced by
$\phi_{\lambda_1}^{\asf\bsf\dagger}
\phi_{\lambda_2}^{\bsf\csf\dagger} \phi_{\lambda_3}^{\csf\asf\dagger}$.
Finally, in place of commutator \rf{27062018-man02-31-b1}, we use
\be
[\phi_\lambda^{\asf\bsf}(p,z),\phi_{\lambda'}^{\asf'\bsf'}(p',z')]\bigl|_{{\rm equal}\, x^+} = \frac{1}{4\beta}\delta^2(p+p')\delta(z-z') \bigl(\delta^{\asf\asf'} \delta^{\bsf\bsf'} + (-)^\lambda \delta^{\asf\bsf'} \delta^{\bsf\asf'}\bigr)\delta_{\lambda+\lambda',0}\,.\quad
\ee
Note also that hermicity of the cubic Hamiltonian leads to the following relations for the coupling constants: $C_\scalarrm^{000*} = C_\scalarrm^{000}$, $C^{\lambda_1\lambda_2\lambda_3*} =  (-)^{\lambda_1+\lambda_2+\lambda_3} \Cb^{-\lambda_1-\lambda_2-\lambda_3}$.

%%%%%%%%%%%%%%%%%%%%%%%%%%%%%%%%%%%%%%%%%%%%%%%%%%%%%%%%%%%%%%%%%%
\newsection{ \large Conclusions}\label{sec-07}
%%%%%%%%%%%%%%%%%%%%%%%%%%%%%%%%%%%%%%%%%%%%%%%%%%%%%%%%%%%%%%%%%%

The light-cone gauge formulation for free fields propagating in AdS space was developed in Ref.\cite{Metsaev:1999ui}. In this paper, we extended the formulation in Ref.\cite{Metsaev:1999ui} to the case of interacting massless fields propagating in $AdS_4$ space. Using such light-cone gauge formulation, we built cubic interaction vertices for arbitrary spin massless fields in $AdS_4$. We found the complete list of such cubic interaction vertices. We expect that our results have the following interesting applications and generalizations.

\noindent \ibf) We built the cubic vertices for light-cone gauge massless AdS fields. Extension of our study to quartic vertices might shed light on our understanding of the locality in the framework  of light-cone gauge formulation of higher-spin field theory. As our cubic vertices for massless AdS fields are similar to the ones for massless fields in flat space \cite{Bengtsson:1986kh}, we expect, in view of results in Refs.\cite{Metsaev:1991mt,Sleight:2016dba}, that the solution for the cubic coupling constants for massless fields in flat space found in Ref.\cite{Metsaev:1991mt} will be valid for massless AdS fields too. Discussion of various methods for analysis of quartic vertices may be found, e.g., in Refs.\cite{Bekaert:2014cea,Dempster:2012vw,Hinterbichler:2017qcl}.

\noindent \iibf) We considered interaction vertices for bosonic AdS fields. It is well known that a supersymmetry leads to additional constraints on interactions vertices. Such constraints might simplify interaction vertices considerably. Therefore, in this respect, it would be interesting to extend our discussion to the case of supersymmetric theories. Recent investigations of various supersymmetric higher-spin theories may be found in Refs.\cite{Kuzenko:2016qwo,Buchbinder:2017nuc,Zinoviev:2007js}. For interesting discussion of arbitrary spin fermionic AdS fields, see Ref.\cite{Najafizadeh:2018cpu}. Recent study of the superparticle in AdS background may be found in Ref.\cite{Uvarov:2018ose}.

\noindent \iiibf) In this paper, we studied fields in $AdS_4$ which, when considering in the framework of Lorentz covariant formulation, are associated with totally symmetric fields of the Lorentz $so(3,1)$. As is well known a string theory involves mixed-symmetry fields. Therefore from the perspective of study of the interrelations between massless higher-spin AdS field theory and string theory it seems important to extend our study to the case of mixed-symmetry fields. Interesting discussion of this theme may be found in Ref.\cite{Sagnotti:2003qa,Vasiliev:2018zer}. Discussion of various interesting Lorentz covariant formulations of free mixed-symmetry fields may be found,  e.g., in Refs.\cite{Campoleoni:2008jq}.%
\footnote{Recent interesting discussion of various group-theoretical aspect of mixed-symmetry (A)dS fields may be found in Refs.\cite{Basile:2016aen}.}
The complete list of cubic vertices for light-cone gauge massless fields in $6d$ flat space was found in Ref.\cite{Metsaev:1993mj} (see also Refs.\cite{Metsaev:2005ar,Metsaev:2007rn}). Particular examples of Lorentz covariant interaction vertices for mixed-symmetry gauge AdS fields  were discussed, e.g., in Ref.\cite{Boulanger:2011se}. Light-cone gauge free mixed-symmetry AdS fields were studied in Refs.\cite{Metsaev:2002vr}-\cite{Metsaev:2015rda}. We believe therefore that the method developed in this paper will allow us to study light-cone gauge interacting mixed-symmetry AdS fields.

\noindent \ivbf) We studied tree-level cubic vertices for arbitrary spin massless AdS fields. Recently the quantum corrections in the theory of higher-spin fields in flat space were studied in Refs.\cite{Ponomarev:2016jqk}. Use of our results for the analysis of quantum corrections for higher-spin AdS fields along the lines in Refs.\cite{Ponomarev:2016jqk} could be very interesting.
We note also that use of our results for the study of AdS/CFT correspondence along the lines in Ref.\cite{Koch:2010cy} might be helpful for better understanding of AdS/CFT correspondence.

\noindent \vbf) Lorentz covariant description of the vertices we obtained in this paper is of some interest.%
\footnote{Along the line in Ref.\cite{Goroff:1983hc}, the  cubic vertex for spin-2 massless field corresponding to the Einstein gravity on AdS background was discussed in Ref.\cite{Akshay:2014pla}.}%
At present time, many promising approaches have been developed for studying Lorentz covariant cubic vertices of AdS fields (see, e.g., Refs.\cite{Vasilev:2011xf,Joung:2011ww}). As we noted, our light-cone gauge vertices for massless AdS fields are closely related to the ones for massless fields in flat space obtained in Ref.\cite{Bengtsson:1986kh}. For the flat space, the discussion of various approaches for studying Lorentz covariant vertices may be found in Refs.\cite{Bekaert:2005jf}-\cite{Henneaux:2012wg}.  However it seems likely that, already for the massless fields in the flat space, covariant description of all light-cone gauge vertices presented in Ref.\cite{Bengtsson:1986kh} is not an easy problem. Recent discussion of this theme may be found in Ref.\cite{Sleight:2016xqq}.

\noindent \vibf) Application of light-cone gauge approach for a study of conformal fields propagating in AdS space and general gravitational background could also be of some interest. Recent studies of Lorentz covariant formulation of conformal fields in general gravitational background may be found in Refs.\cite{Joung:2012qy} . Ordinary derivative formulation of free conformal fields in AdS background was discussed in Ref.\cite{Metsaev:2014iwa}. For recent interesting studies of conformal fields by using twistor-like descriptions, see Refs.\cite{Uvarov:2014lfa}. Discussion of higher-spin conformal fields in the framework of world-line approach may be found in Ref.\cite{Bonezzi:2017mwr}.

\medskip

{\bf Acknowledgments}. This work was supported by the RFBR Grant No.17-02-00546.

%%%%%%%%%%%%%%%%%%%%%%%%%%%%%%%%%%%%%%%%%%%%%%%%%%%%%%%%%%%%%%%%%%%%%%%%%%%%%%%%%%%%%%%%%%%%%%%%%%%%%%%
%%%%%%%%%%%%%%%%%%%%%%%%%%%%%%%%%%%%%%%%%%%%%%%%%%%%%%%%%%%%%%%%%%%%%%%%%%%%%%%%%%%%%%%%%%%%%%%%%%%%%%%
\setcounter{section}{0}\setcounter{subsection}{0}
\appendix{ \large Notation, conventions, and useful formulas}\label{notation}
%%%%%%%%%%%%%%%%%%%%%%%%%%%%%%%%%%%%%%%%%%%%%%%%%%%%%%%%%%%%%%%%%%%%%%%%%%%%%%%%%%%%%%%%%%%%%%%%%%%%%%%
%%%%%%%%%%%%%%%%%%%%%%%%%%%%%%%%%%%%%%%%%%%%%%%%%%%%%%%%%%%%%%%%%%%%%%%%%%%%%%%%%%%%%%%%%%%%%%%%%%%%%%

Throughout this paper, for any quantity $\chi$, the quantity $\partial_\chi$ stands for derivative with respect to $\chi$, while the quantity $N_\chi$ stands for the homogeneity operator,
\be \label{16072018-man02-01}
\partial_\chi \equiv \frac{\partial}{\partial \chi}\,, \qquad N_\chi \equiv \chi \partial_\chi\,.
\ee
We use the following notations for the various quantities constructed out of the momenta $p_a^1$, the radial derivatives $\partial_{z_a}$, the spin operators $M_a^{\Rsm\Lsm}$, and the momenta $\beta_a$, $a=1,2,3$,
\beq
\label{16072018-man02-03} && \Po^1 \equiv \frac{1}{3}\sum_{a=1,2,3} \betach_a p_a^1\,,
\\
\label{16072018-man02-04} && \Pbf_z \equiv \sum_{a=1,2,3}\partial_{z_a}\,,
\\
\label{16072018-man02-05} && \Po_z \equiv \frac{1}{3} \sum_{a=1,2,3}\betach_a \partial_{z_a}\,,
\\
\label{16072018-man02-06} && \PP_z \equiv \sum_{a=1,2,3} \frac{\partial_{z_a}}{\beta_a}\,.
\\
\label{16072018-man02-07} && \Mbf^{\Rsm\Lsm} =  \sum_{a=1,2,3} M_a^{\Rsm\Lsm}\,,
\\
\label{16072018-man02-08} &&  \Mo^{\Rsm\Lsm} =  \frac{1}{3} \sum_{a=1,2,3} \betach_a M_a^{\Rsm\Lsm}\,,
\\
\label{16072018-man02-09} && \MM^{\Rsm\Lsm} =  \sum_{a=1,2,3} \frac{1}{\beta_a} M_a^{\Rsm\Lsm}\,.
\eeq
Various quantities constructed out of the momenta $\beta_1,\beta_2,\beta_3$ and their derivatives
are defined as follows
\beq
\label{16072018-man02-10} && \betach_a \equiv \beta_{a+1} - \beta_{a+2}\,,\qquad \beta_{a+3}= \beta_a\,,
\\
\label{16072018-man02-11} && \Delta_\beta \equiv  \beta_1^2+ \beta_2^2+ \beta_3^2\,,
\\
\label{16072018-man02-12} && \beta \equiv \beta_1 \beta_2\beta_3\,,
\\
\label{16072018-man02-14} && \betach \equiv \betach_1 \betach_2\betach_3\,,
\\
\label{16072018-man02-15} && \No_\beta = \frac{1}{3} \sum_{a=1,2,3}\betach_a \beta_a\partial_{\beta_a}\,.
\eeq
For the momenta $\beta_a$, we have the following helpful relations:
\beq
\label{16072018-man02-16} &&  \sum_{a=1,2,3} \beta_a = 0 \,,
\\
\label{16072018-man02-17} && \sum_{a=1,2,3} \check{\beta}_a= 0\,,
\\
\label{16072018-man02-18} && \sum_{a=1,2,3} \beta_a^2 = \Delta_\beta \,,
\\
\label{16072018-man02-19} && \sum_{a=1,2,3} \betach_a^2= 3\Delta_\beta \,,
\\
\label{16072018-man02-20} && \sum_{a=1,2,3} \frac{1}{\beta_a} = - \frac{\Delta_\beta}{2\beta} \,,
\\
\label{16072018-man02-21} && \sum_{a=1,2,3} \beta_a \betach_a =0 \,,
\\
\label{16072018-man02-22} &&  \sum_{a=1,2,3} \beta_a^3 = 3\beta \,,
\\
\label{16072018-man02-23} && \sum_{a=1,2,3} \betach_a^3 = 3\betach \,,
\\
\label{16072018-man02-24} &&  \sum_{a=1,2,3} \beta_a \betach_a^2 = -9 \beta\,,
\\
\label{16072018-man02-25} &&  \sum_{a=1,2,3} \betach_a \beta_a^2  = - \betach\,,
\\
\label{16072018-man02-26} && \sum_{a=1,2,3}  \frac{\betach_a}{\beta_a} = - \frac{\betach}{\beta} \,.
\eeq
Note that relation \rf{16072018-man02-16} is just the conservation law for the momenta $\beta_1,\beta_2,\beta_3$, while all the remaining relations in \rf{16072018-man02-17}-\rf{16072018-man02-26}, with the exception of \rf{16072018-man02-17},\rf{16072018-man02-18},\rf{16072018-man02-21}, are obtained from \rf{16072018-man02-16}. The relations \rf{16072018-man02-17},\rf{16072018-man02-21} are valid for arbitrary $\beta_1,\beta_2,\beta_3$ in view of definition of $\betach_a$ \rf{16072018-man02-10}, while the relation \rf{16072018-man02-18} is just the definition of $\Delta_\beta$ \rf{16072018-man02-11}.

%%%%%%%%%%%%%%%%%%%%%%%%%%%%%%%%%%%%%%%%%%%%%%%%%%%%%%%%%%%%%%%%%%%%%%%%%%%%%%%%%%%%%%%%%%%%%%%%%%%%%%%
%%%%%%%%%%%%%%%%%%%%%%%%%%%%%%%%%%%%%%%%%%%%%%%%%%%%%%%%%%%%%%%%%%%%%%%%%%%%%%%%%%%%%%%%%%%%%%%%%%%%%%%
\appendix{ \large Derivation of equations \rf{04072018-man02-19}, \rf{04072018-man02-20}}\label{det-01}
%%%%%%%%%%%%%%%%%%%%%%%%%%%%%%%%%%%%%%%%%%%%%%%%%%%%%%%%%%%%%%%%%%%%%%%%%%%%%%%%%%%%%%%%%%%%%%%%%%%%%%%
%%%%%%%%%%%%%%%%%%%%%%%%%%%%%%%%%%%%%%%%%%%%%%%%%%%%%%%%%%%%%%%%%%%%%%%%%%%%%%%%%%%%%%%%%%%%%%%%%%%%%%

In this Appendix, we show that equation \rf{03072018-man02-08} amounts to the two equations given in \rf{04072018-man02-19},\rf{04072018-man02-20}.

First, we derive equation \rf{04072018-man02-19}. To this end, we use the representation for the operators $\Kbf^{1\dagger}$, $\Pbf^-$, and $\Jbf^{-1\dagger}$ given in \rf{04072018-man02-11}, \rf{04072018-man02-12}, and \rf{04072018-man02-14} respectively. Plugging relations \rf{04072018-man02-11}-\rf{04072018-man02-14} into \rf{03072018-man02-08} and considering the zero-order and the first-order terms in $\PP_z$, we get the following two equations for the vertex
\beq
\label{04072018-man02-24} && \hspace{-1cm} \bigl( \Jbf^{z1} + \partial_z \Kbf^{1\dagger}\bigr) |p_\smp3^-\rangle = 0 \,,
\\
\label{04072018-man02-25} && \hspace{-1cm} \Bigl( - \Po^1\No_\beta - \Mo^{z1}\Po_z + \frac{\beta}{3} \MM^{z1} \partial_z +  \frac{\Delta_\beta}{9} \Po_z\partial_{\Po^1}\partial_z +  \frac{\Delta_\beta}{18}  \Jbf^{z1} \partial_z +    \frac{\betach}{54} \partial_{\Po^1} \partial_z^2\Bigr) |p_\smp3^-\rangle
\nonumber\\
&& \hspace{-1.3cm} + \Bigl( \frac{\Po^1\Po^1-\Po_z\Po_z}{2} + \frac{\Delta_\beta}{36} \partial_z^2 \Bigr)\Kbf^{1\dagger} | p_\smp3^- \rangle  = 0\,.
\eeq
Now, plugging $\Kbf^{1\dagger}$ \rf{04072018-man02-11} into \rf{04072018-man02-24}, we recast equation \rf{04072018-man02-24} into the form
\be \label{04072018-man02-26}
\Bigl((N_z + 2) \Jbf^{z1} +  \No_\beta \partial_{\Po^1}\partial_z  - \Mo^{z1}\partial_{\Po_z}\partial_z + \frac{\Delta_\beta}{9}  \partial_{\Po_z} \partial_{\Po^1} \partial^2_z \Bigr) |p_\smp3^-\rangle =  0\,,
\ee
which coincides with the one in \rf{04072018-man02-19}.

Second, we derive equation \rf{04072018-man02-20}. Making use of equation \rf{04072018-man02-24} in the last $\Delta_\beta\partial_z^2$-term in the equation \rf{04072018-man02-25}, we recast the equation \rf{04072018-man02-25} into the following form
\beq
\label{04072018-man02-28} && \hspace{-1cm} \Bigl( - \Po^1\No_\beta - \Mo^{z1}\Po_z + \frac{\beta}{3} \MM^{z1} \partial_z +  \frac{\Delta_\beta}{9} \Po_z\partial_{\Po^1}\partial_z +  \frac{\Delta_\beta}{36}  \Jbf^{z1} \partial_z +    \frac{\betach}{54} \partial_{\Po^1} \partial_z^2\Bigr) | p_\smp3^-\rangle
\nonumber\\
&& \hspace{-1cm} +\,\, \half\bigl(\Po^1\Po^1-\Po_z\Po_z\bigr) \Kbf^{1\dagger} |p_\smp3^- \rangle =   0\,.
\eeq
Multiplying equation \rf{04072018-man02-26} by $(N_z+2)^{-1}$, we get the equation
\be  \label{04072018-man02-27}
\Jbf^{z1}  |p_\smp3^- \rangle =  - \frac{1}{N_z + 2}\Bigl( \No_\beta \partial_{\Po^1}\partial_z  - \Mo^{z1}\partial_{\Po_z}\partial_z + \frac{\Delta_\beta}{9}  \partial_{\Po_z} \partial_{\Po^1} \partial^2_z\Bigr) |p_\smp3^-\rangle\,,
\ee
while plugging \rf{04072018-man02-27} into \rf{04072018-man02-11}, we get the following realization of the operator $\Kbf^{1\dagger}$ on space of $|p_\smp3^-\rangle$:
\be \label{04072018-man02-29}
\Kbf^{1\dagger}  |p_\smp3^- \rangle =   \frac{1}{N_z+1} \Bigl( \No_\beta \partial_{\Po^1}  - \Mo^{z1}\partial_{\Po_z}   + \frac{\Delta_\beta}{9}  \partial_{\Po_z} \partial_{\Po^1} \partial_z \Bigr) |p_\smp3^-\rangle\,.
\ee
Plugging \rf{04072018-man02-29} into the last term in \rf{04072018-man02-28} and using \rf{04072018-man02-09}, we get equation \rf{04072018-man02-20}.

%%%%%%%%%%%%%%%%%%%%%%%%%%%%%%%%%%%%%%%%%%%%%%%%%%%%%%%%%%%%%%%%%%%%%%%%%%%%%%%%%%%%%%%%%%%%%%%%%%%%%%%
%%%%%%%%%%%%%%%%%%%%%%%%%%%%%%%%%%%%%%%%%%%%%%%%%%%%%%%%%%%%%%%%%%%%%%%%%%%%%%%%%%%%%%%%%%%%%%%%%%%%%%%
\appendix{ \large Derivation of relations \rf{15072018-man02-01}-\rf{07072018-man02-19} }
%%%%%%%%%%%%%%%%%%%%%%%%%%%%%%%%%%%%%%%%%%%%%%%%%%%%%%%%%%%%%%%%%%%%%%%%%%%%%%%%%%%%%%%%%%%%%%%%%%%%%%%
%%%%%%%%%%%%%%%%%%%%%%%%%%%%%%%%%%%%%%%%%%%%%%%%%%%%%%%%%%%%%%%%%%%%%%%%%%%%%%%%%%%%%%%%%%%%%%%%%%%%%%

Now we demonstrate that, by using equations \rf{04072018-man02-19},\rf{04072018-man02-20} and \rf{14072018-man02-01}-\rf{14072018-man02-03}, we obtain decoupled equations for the holomorphic and anti-holomorphic vertices \rf{07072018-man02-04}, while for the vertex $V^{000}$ we obtain solution given in \rf{15072018-man02-01}. We split our consideration into four steps.

\noindent {\bf Step 1}. Using the second equations in \rf{14072018-man02-01}-\rf{14072018-man02-03}, we find that dependence of the vertices in \rf{07072018-man02-04} on the momenta $\Po^\Lsm$, $\Po^\Rsm$, and the coordinate $z$ can be presented as
\beq
\label{14072018-man02-05} &&  \hspace{-3cm} V_\scalarrm^{000} =  \frac{1}{z} V_0^{000}\,, \hspace{2cm}   V = \sum_{n=1}^N (\Po^\Lsm)^n z^{n-1} V_n\,, \qquad \Vb = \sum_{n=1}^N (\Po^\Rsm)^n z^{n-1} \Vb_n\,,
\\
\label{14072018-man02-05-a1} &&  \hspace{-3cm}  V_0^{000} =  V_0^{000}(\beta_a,\alpha_a)\,, \hspace{1cm} V_n = V_n(\beta_a,\alpha_a)\,,\hspace{1.7cm}  \Vb_n = \Vb_n(\beta_a,\alpha_a)\,,
\eeq
where, as displayed in \rf{14072018-man02-05-a1}, the vertices $V_0^{000}$, $V_n$, $\Vb_n$ depend only on the momenta $\beta_a$ and the oscillators  $\alpha_a=\alpha_a^\Lsm,\alpha_a^\Rsm$, $a=1,2,3$.
Note also that the first equation in \rf{14072018-man02-01} leads the following equation for the vertex $V_0^{000}$:
\be \label{15072018-man02-02}
\sum_{a=1,2,3}\beta_a\partial_{\beta_a} V_0^{000} = 0\,.
\ee

\noindent {\bf Step 2}. Plugging $p_\smp3^-$ \rf{07072018-man02-03} and \rf{14072018-man02-05} into \rf{04072018-man02-19} and considering $z^{-1}$ term, we get the equation
\be \label{15072018-man02-03}
\Mbf^{z1} | V_0^{000}\rangle = 0 \,, \qquad |V_0^{000}\rangle \equiv V_0^{000}|0\rangle\,,
\ee
while, plugging $p_\smp3^-$ \rf{07072018-man02-03} and \rf{14072018-man02-05} into \rf{04072018-man02-20} and considering  $z^{-1}$ and  $z^{-2}$ terms, we get the respective equations
\beq
\label{15072018-man02-04} && \No_\beta V_0^{000} = 0 \,, \qquad \Mo^{z1} |V_0^{000}\rangle = 0 \,,
\\
\label{15072018-man02-05}  && \MM^{z1} |V_0^{000}\rangle = 0\,,
\eeq
where we use the definitions for the spin operators given in \rf{16072018-man02-08},\rf{16072018-man02-09}.
We now note that equation \rf{15072018-man02-02} and the first equation in \rf{15072018-man02-04} tell us that $V_0^{000}$ is independent of the momenta $\beta_1,\beta_2,\beta_3$.

\noindent {\bf Step 3}. Equations \rf{15072018-man02-03},\rf{15072018-man02-05}, and 2nd equation in \rf{15072018-man02-04} amount to the following three equations
\be \label{15072018-man02-06}
M_a^{z1} |V_0^{000}\rangle = 0 \,, \qquad a=1,2,3\,.
\ee
We note also that ket-vector $|V_0^{000}\rangle$ should satisfy the constraints implied by  \rf{27062018-man02-10},
\be \label{15072018-man02-07}
\alphab_a^\Rsm \alphab_a^\Lsm |V_0^{000}\rangle = 0\,, \qquad a=1,2,3\,.
\ee
From equations \rf{15072018-man02-06}, \rf{15072018-man02-07}, we learn that vertex $V_0^{000}$ is independent of the oscillators $\alpha_a^\Rsm$, $\alpha_a^\Lsm$, $a=1,2,3$. Thus, we conclude that $V_0^{000}$ is a constant which we denote by $C_\scalarrm^{000}$.

\noindent {\bf Step 4}.  Taking into account that $V_0^{000}$ is a constant and using expansions for $V$, $\Vb$ given in \rf{14072018-man02-05}, we verify that equation for $|p_\smp3^-\rangle$  \rf{04072018-man02-19} leads to decoupled equations for holomorphic and anti-holomorphic vertices in \rf{07072018-man02-05},\rf{07072018-man02-06},  while equation for $|p_\smp3^-\rangle$ \rf{04072018-man02-20} leads to decoupled equations for holomorphic and anti-holomorphic vertices  \rf{07072018-man02-18},\rf{07072018-man02-19}.

%%%%%%%%%%%%%%%%%%%%%%%%%%%%%%%%%%%%%%%%%%%%%%%%%%%%%%%%%%%%%%%%%%%%%%%%%%%%%%%%
%%%%%%%%%%%%%%%%%%%%%%%%%%%%%%%%%%%%%%%%%%%%%%%%%%%%%%%%%%%%%%%%%%%%%%%%%%%%%%%%

\end{document}